\documentclass[]{article}
\usepackage[a4paper, total={5.5in, 8in}]{geometry}
\usepackage{graphicx}
\usepackage{color,soul}
\usepackage{cancel}
\usepackage{amsmath}
\usepackage{amsfonts}
\usepackage[mathlines]{lineno}
\usepackage{courier}
\usepackage{multirow}

\newtheorem{assumption}{Assumption}

\newcommand{\interior}[1]{%
	{\kern0pt#1}^{\mathrm{o}}%
}

\definecolor{giu_comm}{rgb}{0.0, 0.42, 0.24}
\DeclareUnicodeCharacter{0308}{HERE!HERE!}

\newcommand{\BibTeX}{{\rm B\kern-.05em{\sc i\kern-.025em b}\kern-.08em\ignorespaces
  T\kern-.1667em\lower.7ex\hbox{E}\kern-.125emX}}

\begin{document}
\overfullrule=9pt
%

\title{\Large{\textbf{Hypoxia-resistance heterogeneity in tumours:\\the impact of geometrical characterization of environmental niches and evolutionary trade-offs.\\A mathematical approach.}}}%
\author{Giulia Chiari$^{*,1,2,3}$, Giada Fiandaca$^{4}$, Marcello Delitala$^{1}$}

\maketitle
\begin{center}
$^{1}$ Department of Mathematical Sciences ``G. L. Lagrange'', Politecnico di Torino (Turin, Italy)\\
$^{2}$ Department of Mathematics ``G. Peano'', Università di Torino (Turin, Italy)\\
$^{3}$ Department of Mathematics, Swinburne University of Technology (Melbourne, Australia)\\
$^{4}$ Department of Cellular, Computational and Integrative Biology, Università degli Studi di Trento (Trento, Italy)\\    
\end{center}
%
%
\begin{abstract}
In the study of cancer evolution and therapeutic strategies, scientific evidence shows that a key dynamics lies in the tumor-environment interaction. In particular, oxygen concentration plays a central role in the determination of the phenotypic heterogeneity of cancer cell populations, whose qualitative and geometric characteristics are predominant factors in the occurrence of relapses and failure of eradication.
We propose a mathematical model able to describe the eco-evolutionary spatial dynamics of tumour cells in their adaptation to hypoxic microenvironments. As a main novelty with respect to the existing literature, we combine a phenotypic indicator reflecting the experimentally-observed metabolic trade-off between the hypoxia-resistance ability and the proliferative potential with a 2d geometric domain, without the constraint of radial symmetry. The model is settled in the mathematical framework of phenotype-structured population dynamics and it is formulated in terms of systems of coupled non-linear integro-differential equations.
The computational outcomes demonstrate that hypoxia-induced selection results in a geometric characterization of phenotypic-defined tumour niches that impact on tumour aggressiveness and invasive ability. Furthermore, results show how the knowledge of environmental characteristics provides a predictive advantage on tumour mass development in terms of size, shape, and composition.
\end{abstract}
\vspace{0.5cm}
\textit{keywords:} continuous structured models, tumour phenotypic heterogeneity,\\tumour-microenvironment interaction, hypoxia, niche construction.\\
\rule{\textwidth}{0.4pt}
\begin{center}
    correspondent author {\texttt{giulia.chiari@polito.it}}
\end{center}

\section*{Introduction}
Cancer is a major cause of mortality throughout the world. According to estimates from the World Health Organization (WHO) in 2019, it was the first or second leading cause of death before the age of 70 years in 112 of 183 countries, and ranks third or fourth in further 23 countries. As a measure of its incidence, in 2020, there were estimated 19.3 million new cases and 10 million cancer deaths worldwide, \cite{sung2021global}. However, despite the extraordinary amount of efforts over the past decades, a successful eradication or control of the advanced forms of this disease remains actually elusive.

In this respect, one of the key aspects of cancer treatments failure is the ability of the malignant masses  to constantly evolve, \cite{ansari2018pancreatic}. This is reflected in the fact that both the genotypic and phenotypic properties of cancer cells may change across space and time within the same tumour and the dynamics of masses with the same histologic features are still likely to vary across patients.
These sources of variability, within and among tumours, provide the substrate for the emergence and development of intra- and inter-tumour heterogeneity which represent major obstacles to cancer eradication, \cite{fisher2013cancer}.

Niche construction theory offers an extraordinary framework to highlight this cancer strength; it identifies indeed in tumour cells the prototypical successful invasive species, able to survive for thousands of generations, constructing niches in which cancer cells acquire added qualities that favour its spreading and its ability to colonize new environments, \cite{ibrahim2017coevolution}.

Oxygen concentration is particularly relevant in this context.
It is indeed clinically observed that, in solid tumours, the oxygen distribution is highly heterogeneous with oxygen levels ranging from normal to mild, almost non-hypoxic, severe hypoxic and even anoxic levels, \cite{nawaz2019analysis}. 
Its tissue deprivation acts as an environmental stressor, promoting a long series of genetic, but especially, epigenetic mutations that strongly impact the tumour eco-evolutionary dynamics. Cancer cells are indeed able to adjust their cellular physiology and metabolism via the up-regulation of different genes as p53, HIF-$\alpha$ or GLUT-1 or IAP-2, acquiring the ability to grow in hypoxic microenvironments and to evade apoptosis, \cite{ruan2009role}.

A key aspect to be considered in this adaptation phenomenon 
is that, as far as talented in adaptation, even cancer cells are subjected to trade-offs in energy allocations to essential functions like growth, maintenance, reproduction and motility. 
Thus, despite their massive evolutionary potential, cancer cells exhibit optimal characteristics along certain traits but are not optimal at all ones, \cite{de2018interplay,ehrlich2017trait}. In this respect, 
it is experimentally observed that cancer cells may be subject to a \textit{trade-off} between \textit{maximizing cell survival} (in terms of an increased tolerance to unfavourable conditions like an oxygen deprived tissue) and \textit{maximizing cell growth}, a concept called \textit{Proliferation-Survival trade-off}. In this respect, it is indeed observed that \textit{hypoxia-resistant} cells experience a decrease in proliferation, showing a doubling time two times longer than normal cells
,\cite{martinez2012hypoxic}.

The role of hypoxia in the eco-evolutionary dynamics of tumour cells is particularly relevant also from a therapeutic point of view. The modalities with which predominant clones appear and go extinct, as well as, if these events take place in bursts or at a more simultaneous pace, require indeed clinical distinctions to determine the more efficient therapeutic approach. 
In this respect, differences in the tumour response could be associated
to the the spatially heterogeneous distribution of intratumoural blood vessels in tumour tissues; their relative dispersal leads indeed to the creation of the above mentioned ecological niches within which tumour cells with different resistance abilities can be selected. Moreover, 
distinct configurations of the blood vessels network could contribute to inter-patients heterogeneity in tumour microenvironment and, consequently, on a disparate tumour response.

Mathematical models could constitute a good investigation instrument in this field to test different environmental conditions, tumour compositions as well as a variety of therapeutic protocols, \cite{clairambault2016physiologically,almeida2018evolution,leszczynski2020optimal}. They can be indeed seen as \textit{in-silico laboratories} to evaluate the consequences of the mutual interactions between the above mentioned aspects on tumour development. In this perspective, additionally, they could accelerate the experimental trials, their theoretical predictions can indeed help to optimize experimental protocols, clinical intervention and limit the number of animal tests, saving money and time during the experimental phases. 

In this respect, cancer cell phenotypic plasticity and its impact on tumours development has been deeply investigated \textit{via} a wide range of modeling techniques. Focusing on continuous models and, in particular, on the mathematical framework of structured populations, different aspects involved in the growth process as resources distribution and/or chemical/physical factors, microenvinronment inhomogeneities and heterogeneity in cancer cells can be seen in \cite{bouin2012invasion,domschke2017structured,jabin2016selection,lorenzi2016tracking}.

In the specific context of hypoxia investigation, integro-differential equations and partial integro-differential have been used in \cite{ardavseva2020mathematical,strobl2020mix,villa2021evolutionary,lorenzi2018role,lorz2015modeling,villa2021modeling, fiandaca2020mathematical} to investigate the ecological role of oxygen distribution in the development of intra-tumour phenotypic heterogeneity.
With a different mathematical approach, the impact of hypoxia on cancer development and on its invasion ability is studied via hybrid cellular automaton models in \cite{gatenby2007cellular}. A recent novel approach is moreover presented in \cite{chiari2022hybrid} where the role of hypoxia, as booster to phenotypic instability that stimulates tumour cells to
shift towards more aggressive hallmarks, is deeply investigated \textit{via} an hybrid approach that characterizes cells both at the genotypic and at the phenotypic level. Moreover, a mechanical model of tumour growth whereby cells switch between aerobic and anaerobic metabolism to survive in hypoxic conditions was presented in \cite{astanin2009mathematical}. 

Motivated from the above considerations, building upon the modeling framework developed in  \cite{fiandaca2020mathematical,villa2021evolutionary,ardavseva2020mathematical}, in this work, we aim to investigate how the mutual interactions between the tumour mass and the resources availability, in particular oxygen distribution, (\textit{i}) can result in a geometric characterization of tumour niches in terms of masses spatial extension and active or necrotic areas formation, (\textit{ii}) how this characterization could affect the phenotypic composition in terms of survival and invasive abilities and finally (\textit{iii}) how both these two aspects in synergy affect the mass growth. Furthermore, this approach lays the groundwork to investigate how the pre-therapeutic history of a tumour, dictated by oxygen distribution, could determine therapeutic failures. The differences in tumour conformation and invasive ability coupled with the emergence of treatment resistant hypoxic cells that result from this dynamics, serve indeed as a nidus for possible subsequent tumour regrowth and repopulation, as well as, for regional and distant dissemination. Thus, they represent
a therapeutic dilemma that needs to be deeper investigated to guarantee the most effective treatment protocol and
to possibly avoid relapses.

The rest of the paper is organized at it follows: in Section \ref{sec:model}, we will present the proposed model with the underlying assumptions. Section \ref{sec:results} will deal with its numerical implementation. In particular, we will first give details on the parameters estimate and on the indices that will quantity tumour progression (see Subsection \ref{sec:simulation_details}). We will then turn on describing the growth of the malignancy in a specific  setting that we will refer as the reference one (Subsection \ref{sec:reference_simulation}). Subsections \ref{sec:results_geomchar} and \ref{sec:results_envsel} will instead be devoted to investigate possible variations with respect to the
reference layout in tumour invasion ability. We will show the model results in a number of selected settings that highlight the role of tumour-microenvironment interaction in tumour development. In particular, in Subsection \ref{sec:results_geomchar}, we will analyse the effect of spatially heterogeneous distributions of the intra-tumoural blood vessels to highlight their role in the creation of ecological niches, due to the relative blood vessels dispersal. Finally, in Subsection \ref{sec:results_envsel}, we will investigate how the strength of the selective pressures exerted by oxygen on tumour cells, in the light of changes in the shape of the proliferation-survival trade-off, impact on the emergence of hypoxic resistance in tumours.  
The article will end in Section \ref{sec:discussion} with a summary of the main results and a discussion on the limitations of our approach with hints for possible developments.

\section{Materials and methods}
\subsection{Mathematical model}
\label{sec:model}

The model here presented will be stated in terms of a system of coupled non linear integro-differential equations; different possible eco-evolutionary scenarios will be explored \textit{via} numerical simulations, varying both the \textit{biophysical / biochemical} cancer cells characteristics and the environmental conditions.
In particular, we will deal with an explicit spatially phenotype-structured population that acquires resources from the environment, provided to the system via a spatial heterogeneous distribution of sources and that diffuse over the environment. Our virtual tumour cells behaviour will be influenced by the phenotypic characteristics of individuals, by the environmental conditions faced and by
the mutual interaction between these two aspects via a spatially-explicit phenotypic dependence. In particular, we will focus uniquely on the role of oxygen that it is here considered as the only available metabolic source.

We assume a tumour mass evolving in a tissue slice i.e. we consider a spatial bi-dimensional domain $\Omega_s \subset \mathbb{R}^2$ in which the tumour mass can expand. 
The population of malignant cells is differentiated in \textit{metabolically active} (\textit{i.e.}, viable) and \textit{necrotic} individuals. The former group is in turn structured with respect to the epigenetic trait $u \in \Omega_p= [0,1]$ that describes their resistance level \textit{i.e.} the ability to survive in harsh environmental conditions, here identified as the hypoxic tumour areas. Specifically, the phenotypic state $u=0$ (\textit{proliferation promoting phenotype}) defines the cell clone with the highest mitotic potential and the lowest level of hypoxia-resistance, whereas $u=1$ (\textit{survival promoting phenotype}) confers the
highest survival ability but the lowest duplication capacity. Between such extreme values, there is a continuum spectrum of possible states that involves the presence of cell variants with intermediate levels of both survival and proliferation. In other words, in a more ecological framework, we are considering a cells population composed by two \textit{specialists}, specifically proliferating and specifically resistant cells, and by a range of \textit{hybrid} ones, the so called \textit{generalists}, able to spend their energies partly to proliferate and partly to survive. 
 This choice to represent the phenotype of a cell via a continuous variable could reflect many biological evidences. As example, it allows to tightly represent how the over-expression of some genes could result in a spectrum of distinct abilities, as shown by the histological data presented in \cite{damaghi2021harsh}, where different levels of expression of GLUT-1 arise in an heterogeneous distribution of cells in terms of hypoxia resistance. In the same veins, it could be particularly suitable to describe the phenotype of a cell as the result of the interaction among several genes that could combine together conferring maxima, minina but also hybrid characteristics, as shown in the Epithelial-Mesenchymal transition, \cite{bhatia2020new}.

According to these hypotheses, the function $a(t,\mathbf{x},u): T \times \Omega_s \times \Omega_p \mapsto \mathbb{R}^{+}_{0}$ hereafter defines the local distribution of active tumour cells on the trait space i.e. $a(t,\mathbf{x},u)$ reflects the phenotypic composition of the tumour mass, located at time $t$ in the domain point $\mathbf{x}$. The local number density of viable individuals can be therefore computed as:
\begin{equation}\label{eq:rho}
	\rho(t,\mathbf{x})=\int_{\Omega_p} a(t,\mathbf{x},u)\, du,
\end{equation}
to consider all the individuals present in the mass, regardless of their phenotype (i.e. integrating on the phenotypic domain $\Omega_p$).

The necrotic subpopulation is instead assumed to be undifferentiated, with number density given by the function $n(t,\mathbf{x}): T \times \Omega_s  \mapsto \mathbb{R}^{+}_{0}$.  
The oxygen concentration is identified by $O(t,\mathbf{x}): T \times \Omega_s \mapsto \mathbb{R}^{+}_{0}$ .

\medskip

\underline{\textit{Cellular scale}}:  Tumour cells proliferate, compete for limited resources, infiltrate the tissue and are subjected to natural selection that optimizes their nutrients uptake (proliferation) and their ability to survive in harsh environments (hypoxia-resistance) taking into account a trade-off that prevents their simultaneous maximization. Cancer cells can also vary their biochemical and biophysical determinants and fall in a necrotic state. Their behaviour is assumed to be affected (according to their phenotypic characteristics) by the surrounding tumour microenvironment, which is, in turn, influenced by the presence of cancer cells in a \textit{feedback/feedforward} fashion. Metabolically active cells are indeed assumed to (\textit{i}) undergo to random phenotypic transitions, (\textit{ii}) randomly move, (\textit{iii}) either proliferate, being subjected to selective pressures by environmental conditions or acquire an irreversible necrotic fate.
The evolution of their distribution can be described by means of the following trait-structured integro-differential equation (IDE):

\begin{equation}
	\label{eq:a}
	\frac{\partial a(t,\mathbf{x},u)}{\partial t} =  \underbrace{\beta_{p} \, \frac{\partial^2 a(t,\mathbf{x},u)}{\partial u^2}}_{\substack{\textup{phenotypic variations}}} +  \underbrace{\beta_s \Delta_\mathbf{x} a(t, \mathbf{x}, u)}_{\textup{movement}}+ \underbrace{R(u,O(t,\mathbf{x}),\rho(t,\mathbf{x}),n(t,\mathbf{x}))a(t,\mathbf{x},u)}_{\textup{proliferation/selection/necrosis}}.
\end{equation}

The diffusion operator at the r.h.s. with respect the $u$ variable of Eq.\eqref{eq:a}, with constant coefficient $\beta_{p}>0$, models infinitesimally small phenotypic variations occurring within the tumour mass as a consequence of random mutation events due to the non-genetic instability that characterizes malignant individuals. This modelling approach is widespread in literature, see for instance the review paper \cite{chisholm2016evolutionary} and references therein, and it is based on the evidence that a cell's phenotype is the result of the interaction between its genetic heritage and the environment in which it lives. In this view, the use of a diffusive term allows to consider two facts: firstly that, in accordance with the biological evidences, the variations in the expression of the genes that could cause small differences in the resulting phenotype, are the most common ones i.e. that are likely to occur with an high probability; secondly that, as observed, there are cases, even if rarely, in which the changes in the expression of only few genes could strongly impact  the observed phenotype, leading to the emergence of individuals with very different characteristics. The diffusion operator at the r.h.s. with respect the $\mathbf{x}$ variable models the  random movement manifested by cells which is described through isotropic Fick's law of diffusion with diffusivity $\beta_s>0$.

The reaction term in Eq.\eqref{eq:a} expresses local variations in the mass of viable cells due to proliferation, the action of the natural selection and necrosis phenomena, whose rates are given by the functions $p$, $S$ and $q$, respectively:
\begin{equation}\label{eq:r}
	R(u,O(t, \mathbf{x}), \rho(t,\mathbf{x}),n(t,\mathbf{x}))=\underbrace{P(u,O(t,\mathbf{x}),\rho(t,\mathbf{x}),n(t,\mathbf{x}))}_{\textup{proliferation}}-\underbrace{S(u,O(t,\mathbf{x}))}_{\textup{selection}}-\underbrace{N(O(t,\mathbf{x}))}_{\textup{necrosis}}.
\end{equation}
In particular, the proliferation rate $p$ is assumed to depend on (\textit{i}) the individual actual phenotype, (\textit{ii}) the resources availability and (\textit{iii}) the physical limitations determined by the available space. In this respect, $p$ is factorized as follows:
\begin{equation}
	P(u,O(t,\mathbf{x}),\rho(t,\mathbf{x}),n(t,\mathbf{x}))=p_1(u) \,p_2(O(t,\mathbf{x})) \, p_3(\rho(t,\mathbf{x}),n(t,\mathbf{x})).
	\label{eq:p}
\end{equation}
The duplication law $p_1$ accounts for the fact that the phenotypic state $u=0$ corresponds to the cell variant with the highest proliferation rate, $\gamma_{\textup{max}}$, whereas a trait value $u=1$ characterizes the cell clone which poorly undergoes mitotic events, as quantified by the lowest rate $\gamma_{\textup{min}}$. In this respect, we work under the hypothesis that the trade-off observed between proliferation and survival is characterized by a linear trend, defining $p_1$ as:
\begin{equation}\label{eq:p1}
	p_1(u)=  (\gamma_{\textup{min}}-\gamma_{\textup{max}})u+\gamma_{\textup{max}}. 
\end{equation}
We are aware that other trade-off shapes could be considered as for instance  the concave one i.e. initially weak and cheap trade-off or the convex one i.e. initially strong and costly
trade-off, \cite{boddy2018life}. To focus on an average condition, we set a linear trade-off.

We then assume that active cells proliferate proportionally to the quantity of oxygen that exceeds a basal concentration $O_{n}$, which corresponds to the amount of molecular substance needed to remain viable and to avoid necrotic transition. More specifically, the relation between cell duplication rate and available chemical is given by a classical Michaelis-Menten law:
\begin{equation}\label{eq:p2}
	p_2(O(t,\mathbf{x}))=\frac{O(t,\mathbf{x})-O_{n}}{\alpha_{\textup{O}}+(O(t,\mathbf{x})-O_{n})} \, H(O(t,\mathbf{x})-O_{n}),
\end{equation}

\noindent being $H(O(t,\mathbf{x})-O_{n})=\left\{1\,,\,\textup{if\,\,}O(t,\mathbf{x})\geq O_{n};\,0\,,\,\textup{if\,\,}O(t,\mathbf{x})< O_{n}\right\}$ the Heaviside function. Eq.(\ref{eq:p2}) therefore implies that mitotic events are prohibited in the case of insufficient presence of oxygen.

The factor $p_3$ in Eq.\eqref{eq:p} finally models the fact that the mitotic cycle is typically disrupted in over-compressed cells. This phenomenon can be replicated by setting the following logistic-like law:
\begin{equation}\label{eq:p3}
	p_3(\rho(t,\mathbf{x}),n(t,\mathbf{x}))= 1- \frac{\rho(t,\mathbf{x})+n(t,\mathbf{x})}{k},
\end{equation}
where $k>0$ is the local tumour tissue carrying capacity. In Eq.(\ref{eq:p3}), we consider that the available space is reduced by the presence of both viable and necrotic individuals.

The function $S(u,O(t,\mathbf{x}))$ in Eq.\eqref{eq:p1} represents the death rate induced by oxygen-driven natural selection. In this respect, starting from the theoretical results and experimental data presented in \cite{korolev2014turning} and \cite{vaupel2001treatment}, we focus on a scenario corresponding to the following biological assumptions:
\begin{assumption}
	\label{oxygenatedenvironment_gg}
	There exist two threshold levels of the oxygen concentration $O_M>O_{m}>0$ such that the environment surrounding the cells is: hypoxic if $O \leq O_{m}$; moderately oxygenated if $O_{m} < O < O_M$;  normoxic (i.e. well oxygenated) if $O \geq O_M$. 
\end{assumption}
\begin{assumption}
	\label{emergenceoffittesttrait_gg}
	The trade-off between the increase in cell death associated with sensitivity to hypoxia and the decrease in cell proliferation associated with acquisition of resistance to hypoxia results in the existence of a level of expression of the hypoxia-resistant gene which is the fittest: a lower level of gene expression would correlate with a lower resistance to hypoxia, and thus a higher death rate; a higher level of gene expression would correlate with a larger fitness cost, and thus  a lower proliferation rate. Cells with levels of gene expression that are closer to the fittest one are more likely to survive than the others. Hence, the farther the gene expression level of a cell is from the fittest one, the more likely is that the cell will die due to a form of oxygen-driven selection.
\end{assumption}
\begin{assumption}
	\label{fittesttrait_gg}
	The fittest level of expression of the hypoxia-resistant gene may vary with the oxygen concentration. In particular: in normoxic environments (i.e. when $O \geq O_M$), the fittest level of gene expression is the minimal one (i.e. $u=0$); in hypoxic environments (i.e. when $O \leq O_{m}$) the fittest level of gene expression is the maximal one (i.e. $u=1$); in moderately-oxygenated environments (i.e. when $O_{m} < O < O_M$), the fittest level of gene expression is a monotonically decreasing function of the oxygen concentration (i.e. it decreases from $u=1$ to $u=0$ when the oxygen concentration increases).
\end{assumption}

Under Assumptions \ref{oxygenatedenvironment_gg}, \ref{emergenceoffittesttrait_gg} and~\ref{fittesttrait_gg}, we define the death rate induced by oxygen-driven selection $S(u,O(t,\mathbf{x}))$ as:
\begin{equation}
	S(u,O(t,\mathbf{x}))=  \eta_O \, \big(u-\varphi_O(O(t,\mathbf{x}))\big)^2,
	\label{eq:selection_gg}
\end{equation}	
where the parameter $\eta_O>0$ is the selection gradient that quantifies the intensity of oxygen-driven selection and the function $\varphi_O(O(t,\mathbf{x}))$ is the fittest level of expression of the hypoxia-resistant gene under the environmental conditions given by the oxygen concentration $O(t,\mathbf{x})$. Its expression, according to Assumption \ref{fittesttrait_gg}, is given by:
\begin{equation}
	\varphi_O(O(t,\mathbf{x}))=
	\begin{cases}
		0, \qquad  \qquad  \qquad \quad  \quad O(t,\mathbf{x})          \geq O_M,\\
		\\
		\displaystyle
		\frac{O_M-O(t,\mathbf{x})}{O_M-O_{m}}, \quad  \quad O_{m} < O(t,\mathbf{x}) < O_M,\\
		\\
		\displaystyle
		1 \qquad  \qquad \quad \qquad \quad O(t,\mathbf{x}) \leq O_{m}.
	\end{cases}	
	\label{phio_So}
\end{equation}

Finally, viable cells irreversibly acquire a necrotic fate when they experience a drop in the available oxygen concentration below to the basal level $O_{n}$. In this respect, the sink term $N$ in Eq.(\ref{eq:r}) reads as follows:
\begin{equation}\label{eq:necr_gg}
	N(O(t,\mathbf{x})) = \eta H(O_{n}-O(t,\mathbf{x})),
\end{equation}
where $H$ is again the Heaviside function and $\eta$ represents a transition rate. We are indeed assuming an inevitable relationship between a low enough amount of resources and a disruption of intracellular metabolic activity, which is in common for all viable cell variants.

\medskip 
\textit{\underline{Necrotic cells dynamics}}: The same rate $N(O(t,\mathbf{x}))$ establishes the growth of the necrotic population as a consequence of the metabolic inactivation of oxygen-deprived cells, i.e.,
\begin{equation} \label{eq:n}
	\frac{\partial n(t,\mathbf{x})}{\partial t}= N(O(t,\mathbf{x})) \rho(t,\mathbf{x}),
\end{equation}
being $\rho(t,\mathbf{x})$, as defined, the number density of viable individuals. Eq.(\ref{eq:n}) implies that necrotic cells remain frozen in space.

\medskip

\textit{\underline{Molecular scale}}:
\noindent  The local concentration of oxygen is governed by a parabolic
PDE where the spatially heterogeneous source term $V(\mathbf{x})$ captures the presence of intra-tumoural blood vessels which bring oxygen into the tumour tissue. Moreover, oxygen diffuses within the tissue, naturally decays and it is consumed by viable cells.\\
Its kinetics is described as follows:

\begin{equation}\label{eq:o}
	\frac{\partial O(t,\mathbf{x})}{\partial t}  =  \underbrace{\beta_{\textup{O}} \Delta_{\mathbf{x}} O(t,\mathbf{x})}_{\textup{diffusion}}-\underbrace{\lambda_{\textup{O}} O(t,\mathbf{x})}_{\textup{natural decay}} - \underbrace{\zeta_{\textup{O}} \int_{\Omega_p} p(O(t,\mathbf{x})) a(t,\mathbf{x},u)\,{\rm d}u}_{\substack{\textup{consumption by}\\ \textup{active tumour cells}}}+\underbrace{V(\mathbf{x})}_{\substack{\textup{inflow from}\\\textup{the blood vessels}}},
\end{equation}

\noindent where $\lambda_{\textup{O}}$, $\beta_{\textup{O}}$ and $\xi_{\textup{O}}$ are constant coefficients.

We denote with ${\Upsilon}=\{(\textbf{v}_i,I_i)\in \Omega_s\times\mathbb{R}^{+}\}_{i=1}^{N_V}$ the set of blood vessels present in the tissue, where the $i$-th vessel is defined by a couple in which the first element $\textbf{v}_i$ provides its the geometrical position and the second element $I_i$ is the rate of inflow of oxygen in the tissue \textit{via} it. Thus, the inflow of oxygen can be described as a geometric source given by:
\begin{equation}
	V(\mathbf{x}) = \sum_{i=1}^{N_V} I_{i} \,e^{- \frac{(\mathbf{x}-{\mathbf v}_i)^2}{\sigma_{V}^2}},
	\label{eq:v}
\end{equation}
with $\sigma_{V}<<1$ to simulate a quasi pointwise source, coherently with the model presented in \cite{villa2021modeling}. In this respect, we specify that, in this modeling arrangement, vessel characteristics, as well as the number of vessels, are time independent. This could be generalized in future, for example considering $N_V=N_V(t)$ to take into account  blood vessel formation or  $I_i=I_i(t)$ to be able to reproduce the effect of external factors, such as therapies, on the oxygen flow provided by the vessels. Moreover, we specify that we do not take into account the effect of mechanical interactions between tumour cells and blood vessels and we do not allow tumour cells to extravasate.
\subsection{Simulation details}
\label{sec:simulation_details}
The spatial domain $\Omega_s$ represents a bi-dimensional section of a 4 cm-large tissue, i.e. $\Omega_s=[-2, 2]^2$ cm. The final observation time, denoted by $t_{\textup{F}}$, varies in each experiment depending on the dynamics to be captured (the longest simulation time-window is $t_{\textup{F}}=1000$ days).

\medskip
\textit{\underline{Initial and boundary conditions}}: Eqs. \eqref{eq:a} and \eqref{eq:n}, that establish cell dynamics, are equipped by the following initial conditions:

\begin{linenomath}
	\begin{equation}
		\label{ica}
		a(0,\mathbf{x},u) = A \ \exp\bigg({-\frac{(\mathbf{x}-\mathbf{x}_C)^2}{2 \sigma_\mathbf{x}^2}-\frac{(u - u_0)^2}{2 \sigma_u^2}}\bigg), \,\, \quad \textup{for}\,\, \mathbf{x},u \, \in \, \Omega_s \times \Omega_p;
	\end{equation}
\end{linenomath}

\begin{linenomath}
	\begin{equation}
		\label{icn}
		n(0,\mathbf{x}) = 0, \quad  \textup{for}\,\, \mathbf{x} \, \in \, \Omega_s,
	\end{equation}
\end{linenomath}
with $A>0 \quad \textup{s.t.}\,\, \rho(0,\mathbf{x})=\int_{\Omega_p} a(0,\mathbf{x},u) du <k$. $\mathbf{x}_C$ is considered as the geometric point around which the cancer cell population is located at the initial time and it is explicitly expressed in every simulation context.

\begin{figure}[htp!]
	\begin{center}
		\includegraphics[width=0.7\textwidth]{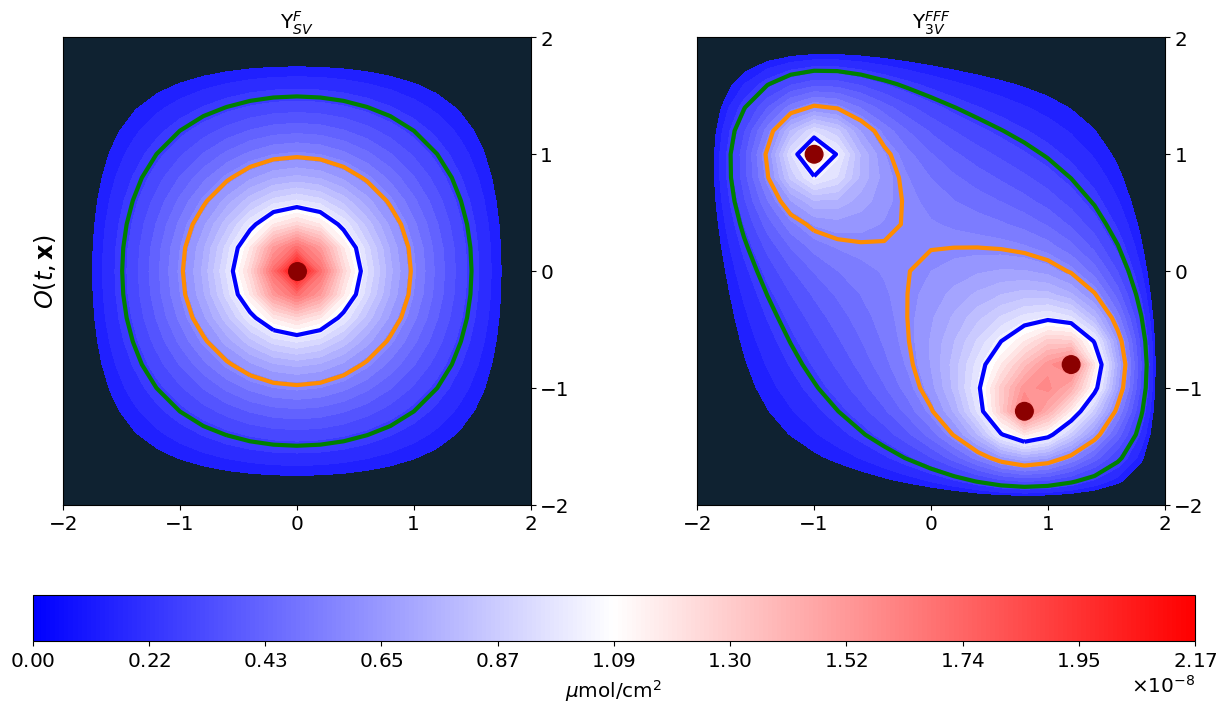}
		\caption{Initial condition for the oxygen map in the two geometrical blood vessels layouts adopted (SV- and 3V-layout, left and right panel respectively). As example, all vessels are set with full intensity (i.e. $\Upsilon^F_{SV}$ and $\Upsilon^{FFF}_{3V}$). The regions outlined by the blue, orange and green lines highlight the optimal areas for low, medium and high phenotypic bands, respectively (the definition of these quantities is provided in \textit{"Quantification of model results"} paragraph in section \ref{sec:simulation_details}).}
		\medskip
		\label{fig:initialconditionoxygen}
	\end{center}
\end{figure}

We are indeed assuming that, at the beginning of all numerical realizations, a node of malignant viable cells is already present within the tissue, with the following characteristics: (i) each cell phenotype has a full Gaussian profile along the spatial dimension, centered at the starting point $\mathbf{x}_C$ and with a variance of $\sigma^2_\mathbf{x}=8 \cdot 10^{-3}$ and (ii) the cell mass has a half-normal distribution in the trait space, with peak at $u_0=0$ being the variance $\sigma^2_u=8 \cdot 10^{-2}$.
This is motivated by the fact that all the simulations begin with a very low number of cells, that is a tumour in an initial phase, and which therefore more easily show the proliferative tendencies needed to settle in the tissue. The initial cell configuration has a maximum value of $A=89.20$ cell/cm$^2$. In this respect, at $t=0$, the overall density $\rho$ of active individuals is symmetrically disposed w.r.t. $\mathbf{x}_C$ and it is mainly composed of proliferative promoting cell variants with only a small fraction of survival promoting agents. At the onset of its growth, the malignancy lacks of a necrotic bulk. 
%
Eq. (\ref{eq:a}) has zero-flux conditions at the boundary of the phenotypic domain, i.e., $\partial_u a(\cdot, \cdot, 0)=\partial_u a(\cdot, \cdot, 1)=0$. This is consistent with the fact that malignant cells can not be characterized by a trait smaller than 0 or higher than 1. The same holds on the domain $\Omega_s$ under the assumption of considering the growth of the mass in a tissue slice where physical barriers as for instance bones, bounds of breast ducts or the lack of extra-cellular matrix, prevent the expansion of the mass out of them.
%
%
Turing on chemical kinetics, Eq. \eqref{eq:o} is completed with initial condition: 
$
O(0,\mathbf{x}) = O^0(\mathbf{x}),
$
 in which, given a blood vessels distribution, $O^0(\mathbf{x})$ represents the the steady-state of oxygen distribution in the tissue in absence of tumour cells. We couple Eq. \eqref{eq:o} with zero-Dirichlet conditions at the boundary of the spatial domain $\Omega_s$ under the assumption of considering a sufficiently large tissue in which anoxic areas are present at the boundaries. 
 In this respect, two geometrical layouts for blood vessels will be adopted in this work. The first and simpler one is composed by a single vessel, placed in the center of the domain $(0,0)$ and we will refer to it as the SV-layout (Single Vessel layout). The second one includes three vessels, placed at points $(-1,1)$, $(1.2,-0.8)$ and $(0.8,-1.2)$, and we will refer to it as the 3V-layout (Three vessels layout). Distinct simulations will take into account the effects of different vessel intensities; in particular, we will denote with $I_{SV}$ and $I_{3V}$ the standard intensities for the SV-layout and the 3V-layout and we will consider different combinations between their full values ($I^F_{SV}$ and $I^F_{3V}$, respectively) and half values ($I^H_{SV}$ and $I^H_{3V}$, respectively). Thus, we can describe all possible configurations of the oxygen sources distribution as: 
\begin{equation*}
	\Upsilon^W_{SV}=\big\{\big((0,0),I^W_{SV}\big)\big\}
\qquad \text{and}
\end{equation*}
\begin{equation}
\Upsilon^{WXY}_{3V}=\big\{\big((-1,1),I^W_{3V}\big),\quad \big((1.2,-0.8),I^X_{3V}\big),\quad \big((0.8,-1.2),I^Y_{3V}\big)\big\},
	\label{eq:layouts}
\end{equation}
with all possible half and full intensity choices $W,X,Y=H,F$, see Figure \ref{fig:initialconditionoxygen} for the disposition of the vessels, as well as the respective oxygen initial condition map, for layouts $\Upsilon^F_{SV}$ and $\Upsilon^{FFF}_{3V}$.

\medskip
\noindent \textit{\underline{Parameters estimate}}:
The majority of model coefficients has a clear and direct biological meaning and therefore a proper estimate has been done taking advantage of the empirical literature. In this respect, we have referred to experimental works dealing with a wide spectrum of diseases since we here account for a generic tumour.

The diffusion coefficient of random phenotypic variations, i.e., $\beta_{p}$, has been taken equal to $8.64 \cdot 10^{-9}$ day$^{-1}$, which is one or two orders of magnitude larger than the rate of somatic DNA mutations, as reported in \cite{doerfler2006dna} in the context of vascularized tumours. The diffusion coefficient in the tissue $\beta_s$ is taken equal to $\beta_s=3.11 \cdot 10^{-5}$ cm$^2$/day, according with the value reported in \cite{martinez2012hypoxic}.
The coefficients $\gamma_{\textup{min}}$ and $\gamma_{\textup{max}}$ quantify the minimal and maximal ability of cells to proliferate according to their phenotypic state $u$ in the case of a given amount of oxygen and available space. The chosen values $\gamma_{\textup{min}}=3.46 \cdot 10^{-1}$  day${}^{-1}$ and $\gamma_{\textup{max}} =6.94\cdot 10^{-1}$  day${}^{-1}$ correspond to a \textit{double time} of 24 and 48 hours respectively, coherently with the biological data reported in \cite{martinez2012hypoxic}. Moreover, they fall within the range of duplication rates evaluated for glioblastoma cell lines in either hypoxic or normoxic conditions, \cite{villa2021modeling}. The carrying capacity has been setting equal to $k=10^6$ cell/cm$^2$ assuming a 10 $\mu$m-mean cell diameter, as computationally measured in \cite{shashni2018size}.  
The concentration of oxygen $O_{n}$, which represent the threshold under which cells go under necrosis, is taken equal to $1.20\cdot 10^{-9}$ $\mu$mol/cm$^2$, while the concentration of oxygen $O_{m}$ that allows cells to remain viable and to duplicate has been taken equal to $2.57\cdot 10^{-9}$ $\mu$mol/cm$^2$. The oxygen concentration threshold $O_{M}=1.37\cdot 10^{-8}$ $\mu$mol/cm$^2$ defines instead the level above which the tissue can be considered in a normoxic condition. These values are taken in agreement with \cite{brown2004exploiting}. The characteristic constant $\alpha_{\textup{O}}$ of the Michaelis-Menten proliferation law has been fixed to $4.28\cdot 10^{-9}$ $\mu$mol/cm$^2$, coherently with the value reported in \cite{dacsu2003theoretical}.
The oxygen diffusion coefficient has been taken equal to $\beta_{\textup{O}}=8.64 \cdot 10^{-1}$ cm$^2$/day, in agreement with the value reported in \cite{martinez2012hypoxic}. 
The oxygen decay rate $\lambda_{\textup{O}} = 8.64 \cdot 10^{-3}$ day$^{-1}$ has been finally set in accordance with \cite{cumsille2015proposal}. All the other parameters are setting model specific coherently with the dynamics of interest.  The full parameters set up is listed in Table \ref{tab1}.

\begin{table}[t!]
	\centering{
		\scriptsize
		\begin{tabular}{lllll}
			\multicolumn{1}{c}{\textbf{}} & \multicolumn{1}{c}{\textbf{Parameter}} & \multicolumn{1}{c}{\textbf{Description}} &  \multicolumn{1}{c}{\textbf{Value [Units]}} & \multicolumn{1}{c}{\textbf{Reference(s)}}\\
			\hline
			\\
			\multirow{10}{*}{\rotatebox{90}{cell dynamics}} & $\beta_{p}$          & phenotypic variation rate     & $8.64\cdot10^{-9}$ [day$^{-1}$] & \cite{doerfler2006dna}   \\ 
			& $\beta_{s}$          &spatial diffusion rate     & $ 3.11 \cdot 10^{-5}$ [cm$^2$/day] & \cite{martinez2012hypoxic}   \\
			
			& $\gamma_{\textup{min}}$         &minimal cell duplication rate     & $3.46\cdot 10^{-1}$ [day$^{-1}$] &\cite{martinez2012hypoxic}  \\
			& $\gamma_{\textup{max}}$         &maximal cell duplication rate     & $6.94\cdot 10^{-1}$  [day$^{-1}$] & \cite{martinez2012hypoxic}  \\
			& $k$         &tissue carrying capacity     & $10^6$ [cell/cm$^2$] & \cite{shashni2018size}  \\
			& $\eta_O$  & oxygen selection gradient & 1 [day$^{-1}$] & model estimate \\
			& $\eta$  & rate of necrotic transition  & 1 [day$^{-1}$] & model estimate \\
			& $A$  & initial maximal cell density   & $89.20$ [cell
			/cm$^2$] & \cite{benzekry2014classical} \\ 
			& $\sigma^2_\mathbf{x}$ & Variance in geometrical space & $0.008$ [cm$^2$] & model estimate \\
			& $\sigma^2_{u}$  & Variance in phenotypic space & $0.08$  & model estimate \\
			\\
			\hline
			\\
			\multirow{10}{*}{\rotatebox{90}{oxygen kinetics}} & $\beta_{\textup{O}}$  & oxygen diffusion coefficient   & $8.64\cdot  10^{-1}$ [cm$^2$/day]  & \cite{martinez2012hypoxic} \\
			& $\lambda_{\textup{O}}$ & oxygen natural decay rate  & $8.64\cdot  10^{-3}$ [day$^{-1}$] &\cite{cumsille2015proposal}  \\
			& $\alpha_{\textup{O}}$  &Michealis-Menten oxygen constant    & $4.28 \cdot 10^{-9}$   [$\mu$mol/ cm$^2$] &\cite{dacsu2003theoretical}  \\
			& $\zeta_{\textup{O}}$ &oxygen consumption rate     & $8.64\cdot 10^{-16}$ [$\mu$mol/cell] &  model estimate  \\
			& $O_{n}$   & oxygen necrotic threshold  & $1.20 \cdot$10$^{-9}$   [$\mu$mol/cm$^2$]  &\cite{brown2004exploiting} \\
			& $O_{m}$   & oxygen hypoxic threshold  & $2.57 \cdot$10$^{-9}$   [$\mu$mol/cm$^2$]   &\cite{brown2004exploiting} \\
			& $O_{M}$   & oxygen normoxic threshold  & $1.37 \cdot$10$^{-8}$   [$\mu$mol/cm$^2$]   &\cite{brown2004exploiting} \\
			& $I^{F}_{SV}$   & full vessel inflow for SV-layout  & $1.58$   [$\mu$mol/cm$^2$ $\cdot$day]   & model estimate \\
			& $I^{F}_{3V}$   & full vessel inflow for 3V-layout  & $1.03$   [$\mu$mol/cm$^2$ $\cdot$day]   & model estimate \\
			\\
			\hline
		\end{tabular}
		\caption{Reference parameters setting.}\label{tab1}}
\end{table}

\medskip
\textit{\underline{Numerical method}}
The domain is discretized as follows:
\begin{itemize}
	\item a uniform one-dimensional discretization is used for the temporal and epigenetic domains;
	\item a triangular mesh with radial symmetry is adopted for the two-dimensional geometric domain.
\end{itemize}
In order to numerically solve the system of partial differential equations, a mixed solution scheme is adopted:
\begin{itemize}
	\item for the one-dimensional components of the domain (time and epigenetic trait), the derivatives are approximated with an explicit Euler method;
	\item for the dynamics on the geometric domain, we apply a Galerkin finite element method, proceeding with a weak formulation of the problem.
\end{itemize}
For the domain mesh and the implementation of the numerical resolution algorithm, a Python code has been developed, using FEniCS and Dolfin packages, \cite{LangtangenLogg2017}.
\subsection{Quantification of model results}
\label{sec:quantification of model results}
As already mentioned, the aim of our work is to investigate how environmental conditions and biophysical determinants of cancer cells influence tumour growth. In this respect, we will focus on the dynamics of observables relative both to the \textit{macroscopic} characteristics of the disease and to its \textit{microscopic} features, i.e., internal heterogeneity. 

To provide some qualitative indicators of tumour evolution and a more quantitative description of phenotypes distribution inside the mass, we divide the epigenetic domain $\Omega_p$ in three, so called, \textit{epigenetic bands}, denoted with L (low), M (medium) and H (high): $\Omega_p=\Omega_p^L\cup\Omega_p^M\cup\Omega_p^H$ with $\Omega_p^L=[0,0.3)$, $\Omega_p^M=[0.3,0.7]$, $\Omega_p^H=(0.7,1]$. 
To enhance the relation between the environmental characteristics and the epigenetic firm of the individuals that colonize them, we link the \textit{epigenetic bands} to the tissue regions in which they are the "optimal" ones \textit{via} the function $\varphi_o(O(t,\mathbf{x}))$ which identifies the fittest trait with respect to the local oxygen concentration $O(t,\mathbf{x})$. In order to do that, we consider separately the area in which oxygen level is below the hypoxic threshold $O_n$ and necrosis occurs; we denote it with $\Omega_{s}^N(t)=\{\mathbf{x}\in\Omega_{s} \, s.t.\,  O(t,\mathbf{x}) \in [0,O_n]\}$ and we will refer to it as the \textit{necrotic area}. Then, we divide the spatial domain coherently with the subsets of the epigenetic domain just presented i.e. in the areas in which the optimal epigenetic traits are, respectively, the ones in the low, medium or high \textit{epigenetic band}, as follows:
$$\Omega_{s}(t)=\Omega_{s}^L(t)\cup\Omega_{s}^M(t)\cup\Omega_{s}^H(t)\cup\Omega_{s}^N(t) \qquad \forall t \in T $$ with $\Omega_{s}^L(t)=\{\mathbf{x}\in\Omega_{s} \, s.t.\,  O(t,\mathbf{x}) \in (\varphi_O^{-1}(0.3),\varphi_O^{-1}(0.0)]\}$, $\Omega_{s}^M(t)=\{\mathbf{x}\in\Omega_{s} \, s.t.\,  O(t,\mathbf{x}) \in [\varphi_O^{-1}(0.7),\varphi_O^{-1}(0.3)]\}$, and  $\Omega_{s}^L(t)=\{\mathbf{x}\in\Omega_{s} \, s.t.\,  O(t,\mathbf{x}) \in [\varphi_O^{-1}(1.0),\varphi_O^{-1}(0.7))\}\setminus O_{s}^N(t)$.
We will refer to $\Omega_{s}^L(t)$, $\Omega_{s}^M(t)$, and $\Omega_{s}^H(t)$ as the low, medium, and high \textit{optimal areas} respectively.

Correspondingly, we introduce their local number densities, denoted with low ($\rho_L$), medium ($\rho_M$) and high ($\rho_H$) respectively, computed as:
\begin{equation}
	\label{eq:rho_I} 
	\rho_I(t,\mathbf{x})=\int_{\Omega^I_p}a(t,\mathbf{x},u)\, du \qquad \text{for } I\in\{L,M,H\}
\end{equation}
and we will refer to them as \textit{band-specific local number densities}. 
Furthermore, in terms of numerosity, we introduce the \textit{total cell count} integrating the local number density on the spatial domain:
\begin{equation}
	\label{eq:cell_count} 
	\Gamma(t)=\int_{\Omega_s}\rho(t,\mathbf{x})\, d\mathbf{x} \end{equation} 
and the corresponding \textit{band-specific cell counts}:
\begin{equation}
	\label{eq:cell_count_band} 
	\Gamma_I(t)=\int_{\Omega^I_s}\rho_I(t,\mathbf{x})\, d\mathbf{x} \qquad \text{for } I\in\{L,M,H\}.
\end{equation}
To have a more clear representation of the phenotypic spectrum present in the mass, we introduce:
\begin{itemize}
	\item the \textit{spatial average epigenetic map}:
	\begin{equation*}\label{eq:spavep}
		f(t,\mathbf{x})=\frac{1}{\rho(t,\mathbf{x})}\int_{\Omega_p}\,a(t,\mathbf{x},u) u\, du,
	\end{equation*}
	that provides an indication of the spatial location of the phenotypic traits;
	\item the \textit{epigenetic global density}:
	\begin{equation}\label{eq:epgldens}
		g(t,u)=\int_{\Omega_{s}} a(t,\mathbf{x},u)\, d\mathbf{x},
	\end{equation} 
	that provides an indication of the number of tumour cells in the entire mass  characterized by a specific epigenetic firm;
	\item the \textit{average epigenetic trait}:
	\begin{equation}\label{eq:aveptr}
		F(t) = \frac{1}{\Gamma(t)} \int_{\Omega_p} u\, g(t,u)\, du, 
	\end{equation} 
	that provides the global average epigenetic trait that characterizes the mass during the evolution.
\end{itemize}
Finally, to evaluate the mass development in terms of spatial extension in the case of tumours characterized by radial symmetry, we introduce the radius measure $r(t)$ which is computed as: 
\begin{equation}
	r(t)= \sup \big\{ \lVert \mathbf{x} - \mathbf{x}_c \rVert \,\,s.t.\,\, \mathbf{x} \in \Omega_s\,\,\text{and} \,\,\rho(t,\mathbf{x})>\sigma_r \big\}
\end{equation}
with $\sigma_r= k/10$ in the meaning of an non-detectable tumour density.
\medskip
\section{Results}
\label{sec:results}
\subsection{Reference simulation}
\label{sec:reference_simulation}
\begin{figure}[htp!]
	\begin{center}
		\includegraphics[width=0.9\textwidth]{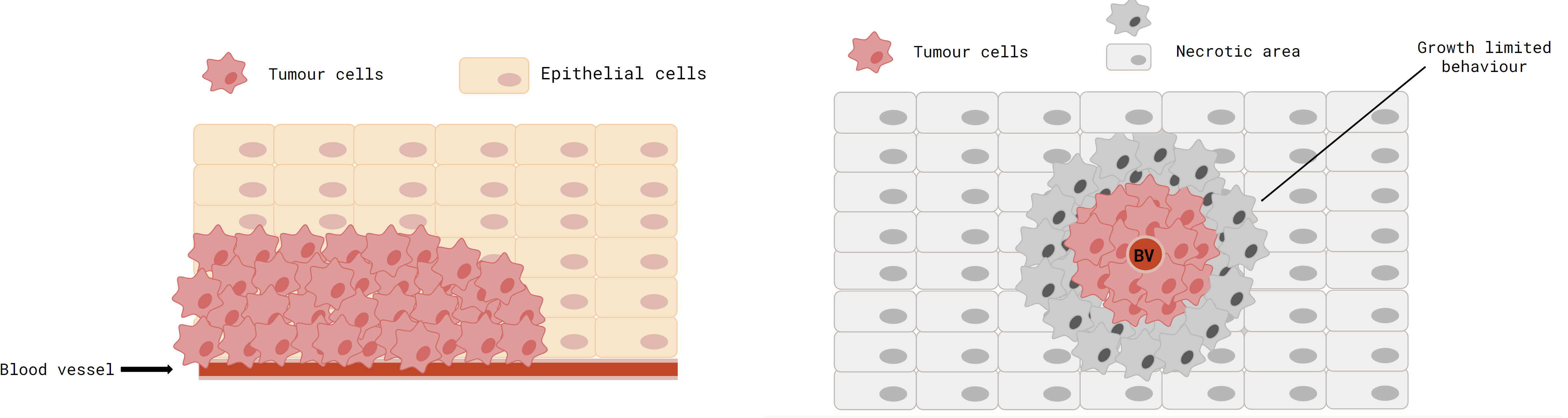}
		\caption{Schematic representation of a tumour cord formation from a cross and a side point of view, (left and right panel, respectively).}
		\label{fig:tumourcord}
	\end{center}
\end{figure}

As prototype of a growing malignant mass in terms of the metabolic switch between normoxic and hypoxic cells, we focus on the investigation of the evolutionary dynamics during tumour expansion in a relatively simple setting. Referring to the possible geometrical layouts for blood vessels cited above, the SV-layout is here settled (see left panel of Figure \ref{fig:initialconditionoxygen}) and, at the beginning of the numerical realization, a single node of malignant viable cells is considered already present at the center of the domain $\mathbf{x}_C=(0,0)$ (in correspondence with the blood vessel) with the phenotypical characteristics defined in Section \ref{sec:simulation_details}. 
The resulting modeling context aims to simulate the development of a \textit{tumour cord} i.e. a cylindrical mass formed by tumour cells that wrap around the blood vessel (see Figure \ref{fig:tumourcord} for a schematic representation of tumour cord formation from a cross and a side view).

\textit{\underline{Oxygen dynamics}}: As it can be observed in the left panel of Figure \ref{fig:initialconditionoxygen}, the reaction-diffusion dynamics of oxygen, along with the inflow through the blood vessel and the consumption by tumour cells,
leads its concentration to converge to stable values which decrease moving away from the source. In particular, its gaussian-like profile appears to be, at numerical equilibrium, as a monotonically decreasing function with respect the distance from the blood vessel, geometrically characterizing the tumour microenvironment with oxygen levels that range from normoxic to mild, severe and even anoxic. 

\begin{figure}[htp!]
	\begin{center}
		\includegraphics[width=0.85\textwidth]{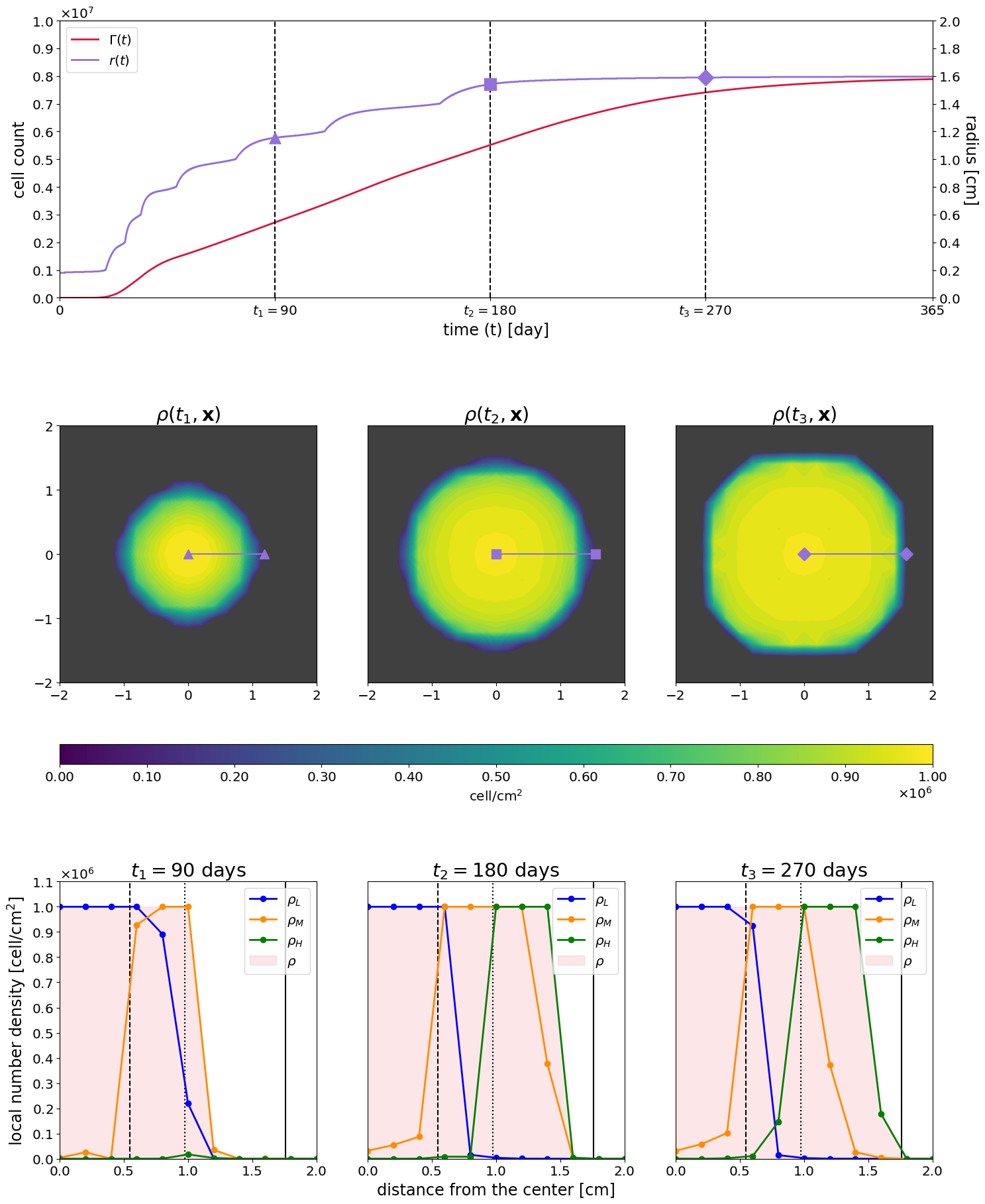}
		\caption{Results of the reference simulation relative to cancer cells dynamics and tumour morphology. (a) First row represents the evolution in time of the total cell count $\Gamma(t)$ and radius $r(t)$ of the tumour mass. Vertical lines detect the times chosen for instantaneous representation in second and third rows. (b) Second row is a two-dimension spatial representation of $\rho(t,\textbf{x})$ for $t=90,180,270$ days. The cancer mass radius, along which plots in the third row are projected, is highlighted with the lilac segment. (c) Third row provides a one-dimension representation of \textit{band-specific} and \textit{global number densities} $\rho_L(t,\textbf{x})$, $\rho_M(t,\textbf{x})$, $\rho_H(t,\textbf{x})$, $\rho(t,\textbf{x})$ along the segment $\overline{(0,0),(2,0)}$ for $t=90,180,270$ days.}
		\label{fig:reference_morpho}
	\end{center}
\end{figure}

\textit{\underline{Cell dynamics}}:
From the morphological point of view, the global behaviour of our virtual mass can be described observing its evolution in terms of spatial extension \textit{via} its radius measure $r(t)$ (top panel of Figure \ref{fig:reference_morpho} - lilac line) and volume \textit{via} its total cells count $\Gamma(t)$ (top panel of Figure \ref{fig:reference_morpho} - red line). As it can be detected in their time evolution, after a first plateau phase, an increment, characterized by a velocity that steadily decreases, is observed during the entire monitoring time-window.

In particular, the total number of active individuals $\Gamma(t)$ in the earliest phases shows an exponential-like growth that quickly shifts on a semi-linear trend and finally approaches a steady-state value ($t \approx 365$ days) in a second phase with respect to the time at which the tumour cord reaches its maximum expansion ($t \approx 270$ days), reproducing an invasion dynamics coherent with the physical constraints. The initial fast growth dynamics observed, in terms of both numerousness and radial expansion, is indeed in line with the mass development in regions which are rich of resources and mainly composed by \textit{proliferation promoting} cells. 
The subsequent slower dynamics is instead observed when the mass approaches increasingly harsh regions with low nutrients concentration that promote the expansion of more resistant phenotypes 
characterized by a lower mitotic potential. Consequently, the radial expansion of the mass, as well as the increment in numerosity, is slower since cells need more time to conquer and to populate new regions.

\begin{figure}[htp!]
	\begin{center}
		\includegraphics[width=0.85\textwidth]{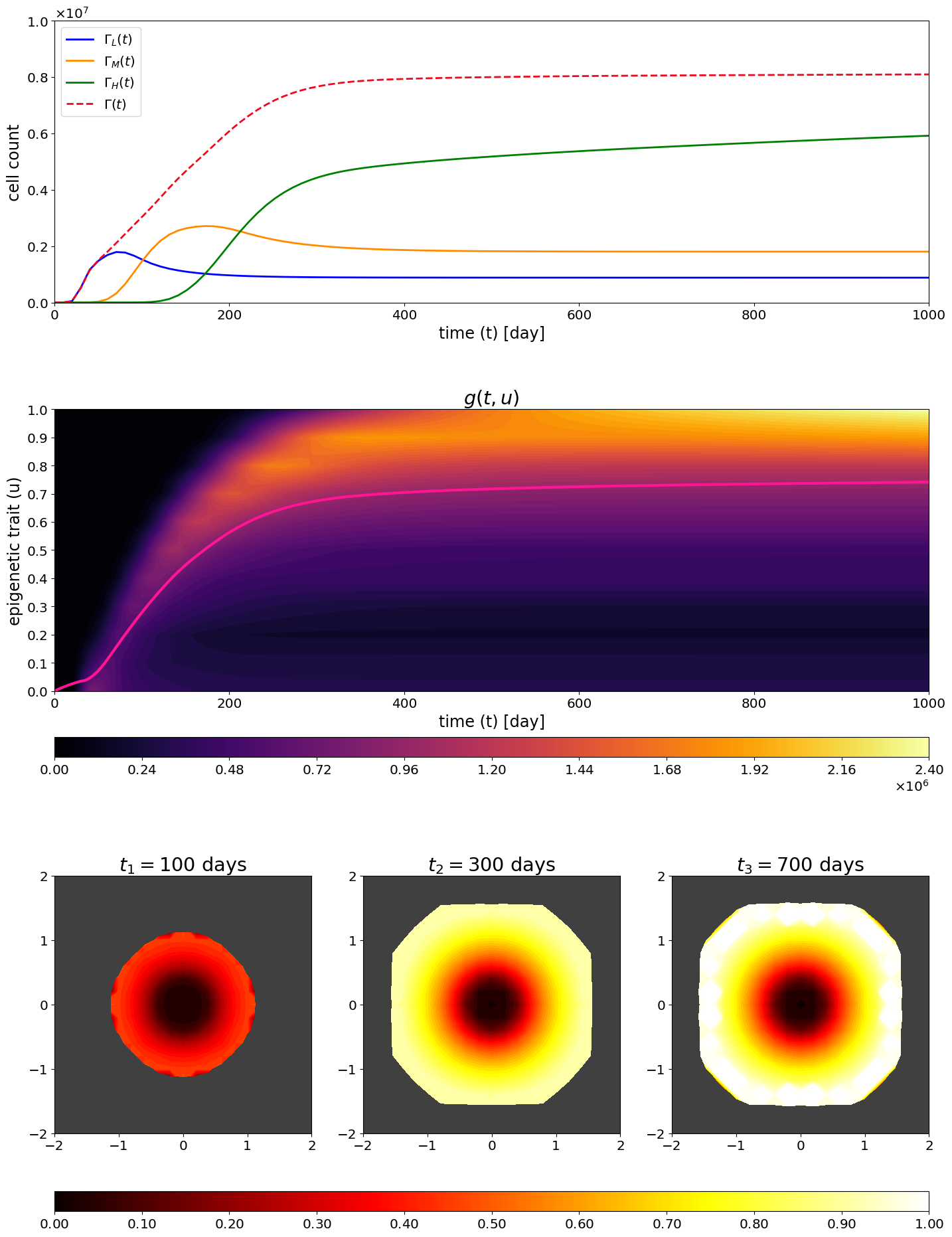}
		\caption{Results of the reference simulation relative to numerical and geometrical phenotypic composition. (a) First row represents the evolution in time of band specific and global cell counts $\Gamma(t)$, $\Gamma_L(t)$, $\Gamma_M(t)$, $\Gamma_H(t)$. (b) Second row shows the epigenetic global density $g(t,u)$. Pink line represents the time evolution of the average epigenetic trait $F(t)$. (c) Third row provides the spatial two-dimension representation of the average epigenetic map $f(t,\mathbf{x})$ at times $t=t_1=100$ days, $t=t_2=300$ days, $t=t_3=700$ days. Definition of these quantities is provided in \textit{"Quantification of model results"} paragraph in section \ref{sec:simulation_details}.}\medskip
		\label{fig:reference_phenotypiccomposition}
	\end{center}
\end{figure}

A multi-phase growth of this type recalls the data reported in different experimental works from the early 70s in which the self-regulation of growth of three dimensional spheroids is studied, \cite{folkman1973self}, and in more recent works in which the volume extensions of different cell lines cultured in vitro is analysed, \cite{oraiopoulou2017vitro}. In this respect, a Gompertz-like kinetics can be observed extending the monitoring time window (results shown in top panel of Figure \ref{fig:reference_phenotypiccomposition}); the cells total count $\Gamma(t)$ shows indeed a sigmoidal  profile that asymptotically converges to a maximal threshold value which corresponds to the maximum carrying capacity of the entire tissue.


In the middle panel of Figure \ref{fig:reference_morpho}, the evolution of the virtual mass is shown in terms of its cell number density $\rho(t,\mathbf{x})$ at three different time ($90$, $180$ and $270$ days). Cell clones start in fact to radially spread, moving away from the blood vessel and colonizing all the available space until the nutrients concentration is sufficient to guarantee their survival. When they reach anoxic tissue regions, cells face a necrotic transition, leading to a cutoff in the invasion dynamics. This growth limited behaviour is coherent with the evolutionary dynamics observed during \textit{in vivo} neoplasms development. Experimental data show indeed that, in avascular phase, neoplasms grow up to a limit average radius and are surrounded by necrotic regions; this quasi-steady state is reached because of the high oxygen consumption by proliferating cancer cells coupled with the diffusion limits of oxygen that lead to a radial decay of the essential nutrients and rise to anoxic areas formation.

Observing the evolution of the mass from a radial cross sectional point of view, shown in the bottom panel of Figure \ref{fig:reference_morpho}, we can better appreciate how the synergistic interaction between cell proliferation and cell movement allows tumour cells to invade the surrounding tissue. The cell number density $\rho(t,\mathbf{x})$, along the segment $\overline{(0,0),(2,0)}$, behaves indeed like an invading front whereby growth is saturated at the local carrying capacity of the tissue $k$, as shown by the pink coloured regions that summarise its evolution at three different times points, ($t=90, 180,270$ days).

In this respect, our findings moreover show how tissue colonization results from the cooperative relations between different specialized cell variants, enhancing the importance of phenotypic composition on tumour development. Analysing indeed the \textit{bands specific number density} of the different sub-groups $\rho_I$ for $I=L,M,H$ (low, medium and high hypoxia-resistant cells - blue, orange and green curves, respectively), it can be noticed that the tumour composition dynamically changes mutually shaping with tumour microenvironment. The spatial variability of oxygen concentration leads to the formation of environmental gradients resulting in the selection for cells with phenotypic characteristics that vary with distance from the blood vessel.
The emergence of specific phenotypic traits is observed in relation to the region of the tumour mass analysed. Black vertical lines distinguish the oxygenation areas: the dashed one divides the optimal areas for low and medium band, the dotted one divides the optimal areas for medium and high band, and the continuous one indicates the limit of the non-necrotic area. In particular, as it can be observed, the more popular clones in each region are promptly characterized by a balance between the ability to duplicate and the decrease in proliferation rate due to higher survival abilities that guarantees them to be the more adapted with respect the environmental conditions faced.

Coherently with the biology behind, this selection mechanism, as it can be noticed comparing left and central panel in the bottom row of Figure \ref{fig:reference_morpho}, is not instantaneous. The time evolution of the different cell fractions reveals indeed that, in a first phase (Figure \ref{fig:reference_morpho}, bottom panel, left side), \textit{proliferation promoting} cells are the predominant clones (blue curve) able to colonize also more distant regions from the blood vessel. In the subsequent phases (Figure \ref{fig:reference_morpho}, bottom panel, central and right sides), we can instead observe the progressive disappearance of these cell variants in the more external regions of the mass coupled with the rising of new clones characterized by an increasingly over-expression of \textit{survival promoting} genes (orange and green curves). At the end of the simulation time, a ring structure can be observed constituted by a central group of proliferating cells surrounded by two concentric rims of medium and high resistant cells, reproducing the evolutionary selection of more resistant cell clones in the case of harsh tissue conditions (and vice versa) whose examples can be been found in literature. In this perspective, our results are not only able to capture this phenomenon but also to give hints on the modalities and the timing at which it takes place. 

Enlarging the time window of observation to better appreciate the evolutionary phenomena that need a longer time-scale to be detected, we can observe that, at the level of the entire mass, a trend to hypoxia-resistance development characterizes the tumour evolution, as shown in Figure \ref{fig:reference_phenotypiccomposition}. 
A gradual phenotypic shift of the entire disease towards more resistant to hypoxia cell phenotypes is indeed observed looking at the evolution in time of the \textit{band-specific cell counts} $\Gamma_I(t) \,\, \text{for}\,\,I \in {L,M,H}$ across the entire mass, as represented in top panel of Figure \ref{fig:reference_phenotypiccomposition}.
Globally indeed, in the first phases of evolution, the mass is mainly composed by \textit{promoting proliferation} cells (blue curve) that in time are out-competed by more and more resistant phenotypes (medium and high resistant cells - orange and green lines respectively). At the final time of observation ($t=1000$ days), high resistant cells constitute approximately the $70$\% of the mass, almost a $20$\% are medium resistant cells and only a small portion are the low resistant ones.

This is coherent with the evolution of the \textit{epigenetic global density} $g(t,u)$, represented in middle panel of Figure \ref{fig:reference_phenotypiccomposition}, which indeed concentrates around $u=0$ in the first phases and gradually shifts towards high resistance development revealing \textit{via} its shape informations on how the emergence of intra-tumour heterogeneity during the growth of the mass takes place. As it can be observed, the spectrum of the phenotypes observed at the final simulation time ($t=1000$ days) is widely expanded with respect the first phases where only \textit{proliferation promoting} cells appear, covering all the phenotypes from the lowest to the higher resistant ones with an increasing intensity towards resistance development, coherently with the results shown in Figure \ref{fig:reference_morpho} and \ref{fig:reference_phenotypiccomposition} and with the evolution of the \textit{average epigenetic trait} $F(t)$ of the mass (pink line).

This is confirmed analysing the \textit{spatial average epigenetic map} $f(t,\mathbf{x})$, bottom panel of Figure \ref{fig:reference_phenotypiccomposition}, which evolution is showed at three different time steps, ($t=100, 300, 700$ days). It reflects the earlier described dynamics in the analysis of the cross section mass development in terms of its phenotypic composition on a longer time-window, coherent with the evolutionary time-scale. In particular, the ring structure previously described is again vividly  detected, moreover revealing that, at the end of the mass expansion, the more popular epigenetic firms are the ones between $u=0.7$ and $u=1$ i.e. the more resistant phenotypes. 
 
Summarizing, our results suggest that, in the early stages of progression, tumour growth and expansion within the host rely on collective cell dynamics which is facilitated by the emergence of an intratumoural phenotypic heterogeneity, showing how cells with different characteristics and functions can cooperate to survive and efficiently invade the host, as it will be deeply investigated in the next sections.
\subsection{The impact of geometric characterization of the environment on tumour niches}
\label{sec:results_geomchar}
To investigate the relation between a tumour and the environment in which it develops and, in particular, the role of hypoxia in this dynamics, the study of the vascular network, responsible for tissue oxygenation, plays a central role.
Different layouts of blood vessels, which vary in inflow intensity and geometric arrangement, can indeed naturally lead to distinct spatially heterogeneous distributions of oxygen that, coupled with  various primary tumour onset in the tissue, could completely change the invasion ability of a mass.
\begin{figure}[htp!]
	\begin{center}
		\includegraphics[width=0.85\textwidth]{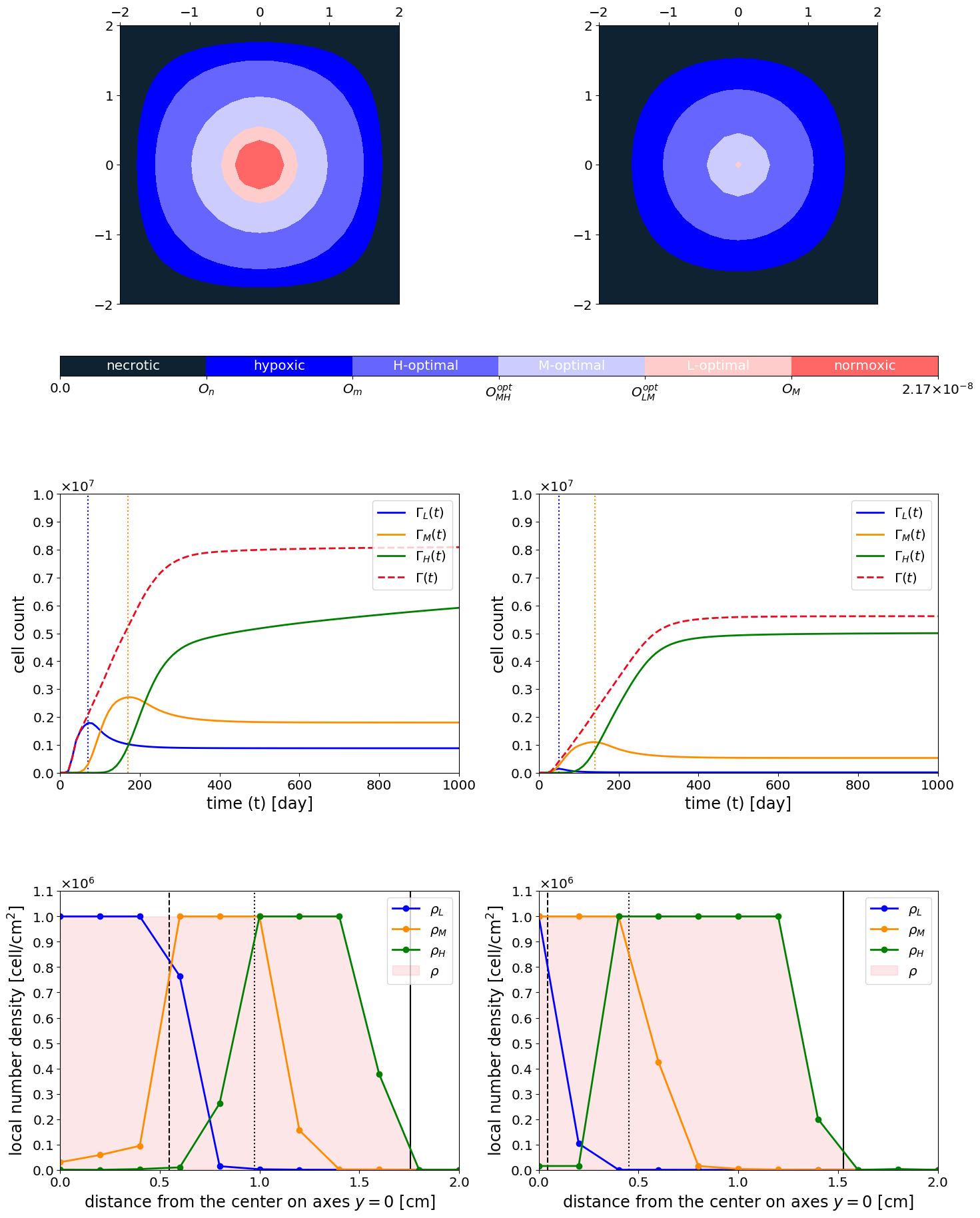}
		\caption{This experiment compares results from the simulation in the reference setting ($\Upsilon^F_{SV}$, left column), with results from a simulation with same parameters and setting with the exception of and half value for the vessel intensity ($\Upsilon^H_{SV}$, right column). (a) First row shows the initial oxygen map $O(t_0,\mathbf{x})$. (b) Second row provides evolution in time of \textit{band-specific} and \textit{global cell counts} $\Gamma(t)$, $\Gamma_L(t)$, $\Gamma_M(t)$, $\Gamma_H(t)$. (c) Third row represents the profiles of the densities $\rho $, $ \rho_L(t,\mathbf{x}) $, $ \rho_M(t,\mathbf{x})$, $ \rho_H(t,\mathbf{x})$, along the segment $\overline{(0,0),(2,0)}$ at the final observation time $t=t_F=1000$ days. Vertical lines partition the spatial domain, from left to right, in $\Omega_s^L(t_{\textup{F}})$, $\Omega_s^M(t_{\textup{F}})$, $\Omega_s^H(t_{\textup{F}})$, and $\Omega_s^N(t_{\textup{F}})$ i.e. the final low, medium, and high optimal areas and the necrotic one.}
		\label{fig:intensity_sv}
	\end{center}
\end{figure}

In this light, in the fist part of this section, we focus on investigating the impact of oxygen inflows intensity on tumour morphology in terms of growth rate, size and epigenetic composition. Figure \ref{fig:intensity_sv} refers to the same set up of the reference simulation, with the only exception of the vessel intensity. The experiment is indeed repeated twice: the graphs in the left column refer to a tumour cord that develops around a vessel with the same intensity as in the reference case ($\Upsilon^F_{SV}$), while those in the right column consider a vessel with half intensity ($\Upsilon^H_{SV}$). 

The first row represents the initial oxygen distribution in the two cases. To enhance the relation between the environmental characteristics and the epigenetic firm of the individuals that
potentially will colonize them, indicative bands, relative to its concentration levels, are represented. In particular, the \textit{anoxic} area in which oxygen drops below the $ O_n $ threshold leading to tumour cells necrosis, is highlighted in dark blue; the \textit{hypoxic} area, characterized by oxygen levels between $ O_n $ and $ O_m $, in which the fittest level of gene expression is the maximal one (i.e. $u=1$), is represented in blue; the \textit{moderately-oxygenated} area, in which oxygen levels vary from $ O_m $ and $ O_M $, is then divided in the three different bands, $\Omega_s^H$, $\Omega_s^M$ and $\Omega_s^L$, relative to the optimal survival areas of high (indaco), medium (light blue), and low (light pink) \textit{epigenetic bands}, $\Omega_p^H$, $\Omega_p^M$ and $\Omega_p^L$ respectively; finally, the \textit{normoxic} area, characterized by oxygen levels higher than $O_M$, in which the fittest level of gene expression is the minimal one (i.e. $u=0$), is represented in cherry.

Comparing the two oxygen maps, differences in proliferation speed, extension and epigenetic distribution of the tumour mass can be predicted. In fact, a faster growth of the mass in the case of a higher oxygen concentration (left column of Figure \ref{fig:intensity_sv}) is observed due to both a more efficient oxygenation on proliferation 
and the presence of a substantial node of \textit{proliferation-promoting} individuals. In contrast, in the case of a more hypoxic tissue, higher epigenetic traits will be selected by the environment, resulting in a tumour with slower growth but with a higher hypoxia-resistance (right column of Figure \ref{fig:intensity_sv}). 

The trend in hypoxia-resistance boost, detected in both cases, is reflected in the evolution in time of the \textit{total} and the \textit{band-specific cell counts}, represented in the second row of Figure \ref{fig:intensity_sv} where vertical blue and yellow lines indicate the time instant at which $\Gamma_L(t)$ and $\Gamma_M(t)$ reach their maximum value, respectively. Comparing the two dynamics, we can see that in the case of a lower oxygenated environment, the \textit{global cells count} $\Gamma(t)$ (magenta dotted line - right panel) grows slower than the corresponding one in the higher oxygenation case (magenta dotted line - left panel) and a lower plateau value is moreover observed in accordance with the presence of 
lower nutrients source.

Notably, in both the two experiments, the \textit{band-specific cell counts} ($\Gamma_L(t)$, $\Gamma_M(t)$, $\Gamma_H(t)$, blue, orange and green curves respectively) highlight the evolutionary concept of inter-competition between species and the so called \textit{competitive exclusion principle} that states that two species with similar needs cannot exist sympatrically, one will always out-compete the other which will either adapt or be excluded, by either emigration or extinction. The higher proliferative rate gives indeed an initial evolutionary advantage to \textit{proliferation-promoting} phenotypes on the other epigenetic traits that overcome in number in the population. Subsequently, in the areas of high oxygenation where the low epigenetic trait is optimal, this dominance is preserved; where instead the oxygenation is lower, the selective pressure by the environment induces a higher mortality rate of low epigenetic traits. Thus, \textit{proliferation-promoting} cells quickly face apoptosis, freeing up space to expand to other cancer subpopulations that, despite the slower proliferation, have a greater resistance to the hostility of the environment leading to a out-competition-like dynamics.

Finally, the third row of Figure \ref{fig:intensity_sv} represents the cross-profiles of the densities $\rho $, $ \rho_L $, $ \rho_M $, $ \rho_H $, along the segment $\overline{(0,0),(2,0)}$ at the final observation time $t_F=1000$ days. The dynamics of the phenotypic diversity observed in the tissue areas colonization, follows  a similar pattern as previously described on the in the corresponding panel of Figure \ref{fig:reference_morpho},  related to the case of a more hypoxic environment.
As expected, both high resistant phenotypes optimal area $\Omega_s^H$ and the necrotic region $\Omega_N$ show a greater extension coherently with a less efficient vasculature, leading to a less extended mass as it can be detected via the radius measure which results to be roughly about $20\%$ smaller than the one in reference case ($r(t_F) \approx 1.5$ cm \textit{vs} $r(t_F) \approx 1.8$ cm).
Summarizing, our findings suggest that tumour success in terms of adaptation, survival and expansion is closely related to the niche characteristics that tumour cells face, coupled with the epigenetic firm optimality with respect to them.

We now turn on to test how much the morphology of the mass as well as its phenotypic composition could change in a wide range of oxygen maps. 
To do this, we fix the tumour primary nidus at the center of the domain and we consider the three vessels layout above described ($3$V-layout), varying the blood vessels intensity to comprehend all the possible combinations of full-intensity and half-intensity inflows.

\begin{figure}[htp!]
	\begin{center}
		\includegraphics[width=0.85\textwidth]{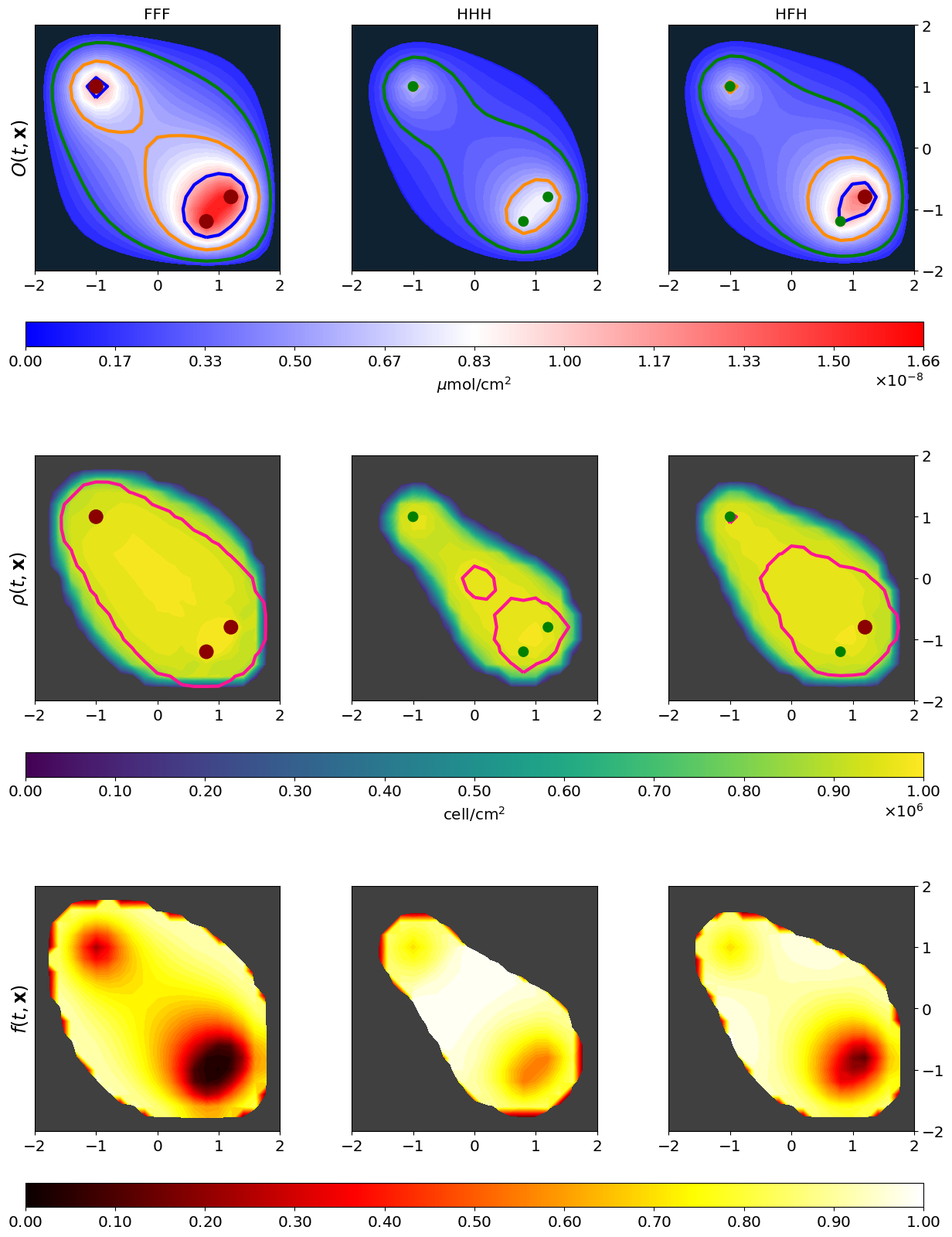}
		\caption{This experiment compares the results from three different simulations in which we vary the blood vessels network in terms of potency. In particular, we adopt the 3V-layout, keeping unaltered parameters from the reference simulation, with the exception of the intensities of the vessels. All possible combinations between full (F) and half (H) intensities are considered. In this plot we show results for the configurations: FFF (first column), HHH (second column) and HFH (third column). (a) First row shows the initial oxygen maps. (b) In second row, pink lines identify the tumour edge at an intermediate time $t_1=100$ days, while the contour plot represents the $\rho(t, \mathbf{x})$ at the final time $t=t_2=200$ days. (c) Third row shows $f(t,\mathbf{x})$ at the final time $t_2=200$ days.}
		\label{fig:multvessels}
	\end{center}
\end{figure}

In Figure \ref{fig:multvessels}, we provide the results of three different settings: (\textit{i}) the one in which all full-intensity vessels are considered ($\Upsilon^{FFF}_{3V}$ configuration, results shown in the first column), (\textit{ii}) the one in which all half-intensity vessels are considered ($\Upsilon^{HHH}_{3V}$ configuration, results shown in the second column) and (\textit{iii}) a third mixed setting in which two vessels characterized by half-intensity and one vessel by the full one are considered (specifically, $\Upsilon^{HFH}$ configuration, results shown in the third column).

First row of Figure \ref{fig:multvessels} shows the  oxygen maps in which, case by case, we indicated the boundaries of corresponding optimal areas $\Omega_s^L$, $\Omega_s^M$ and $\Omega_s^H$ for low $\Omega_p^L$, medium $\Omega_p^M$ and high $\Omega_p^H$ \textit{epigenetic bands} with the same colors used for the \textit{band-specific cell counts} (blue, orange and green-coloured contours, respectively). 
Second row of Figure \ref{fig:multvessels} provides the tumour local density profile $\rho$ at the final time of observation $t_2=200$ days on which the boundary of the tumour mass at an intermediate time $t_1=100$ days is projected in pink. 
{Finally, third row of Figure \ref{fig:multvessels} represents the different epigenetic distributions of tumour cells in each case. Comparing them to the oxygenation maps reported in the first row, we can observe that in the first case (first column), the pretty extended optimal area $\Omega_s^L$ for the \textit{low-specific epigenetic band} $\Omega_p^L$ , is reflected by the presence of the two dark spots in the corresponding areas of the graph in the third row. In the same veins, the graphs in the third and, even more evidently, in the second column have lighter colors, a fact that recalls the predominance of optimal areas for medium and high bands.} 

{Notably, our findings suggest that a tumour mass which develops in well oxygenated area tends to assume a profile close to its steady state earlier than the one that grows up in an adverse environment as it can be observed looking at the pink boundaries projected in the second row. In our perspective, these results could be particularly interesting from a medical perspective in the cases in which a  surgical removal of a mass has to be planned. They are indeed able to suggest in which cases and in which terms a larger region with respect the one detected via imagine analysis at some point could be more safer to be removed. Specifically, they indeed show that in the case of a harsh microenvironments, like the one represented in the second column, an higher probability of the presence of silent regions in terms of density but characterized by a strong invasive ability in which the mass is already growing has to be taken into account.}

\subsection{The impact of environmental selection forces}
\label{sec:results_envsel}
\medskip
\begin{figure}[htp!]
	\begin{center}
		\includegraphics[width=0.82\textwidth]{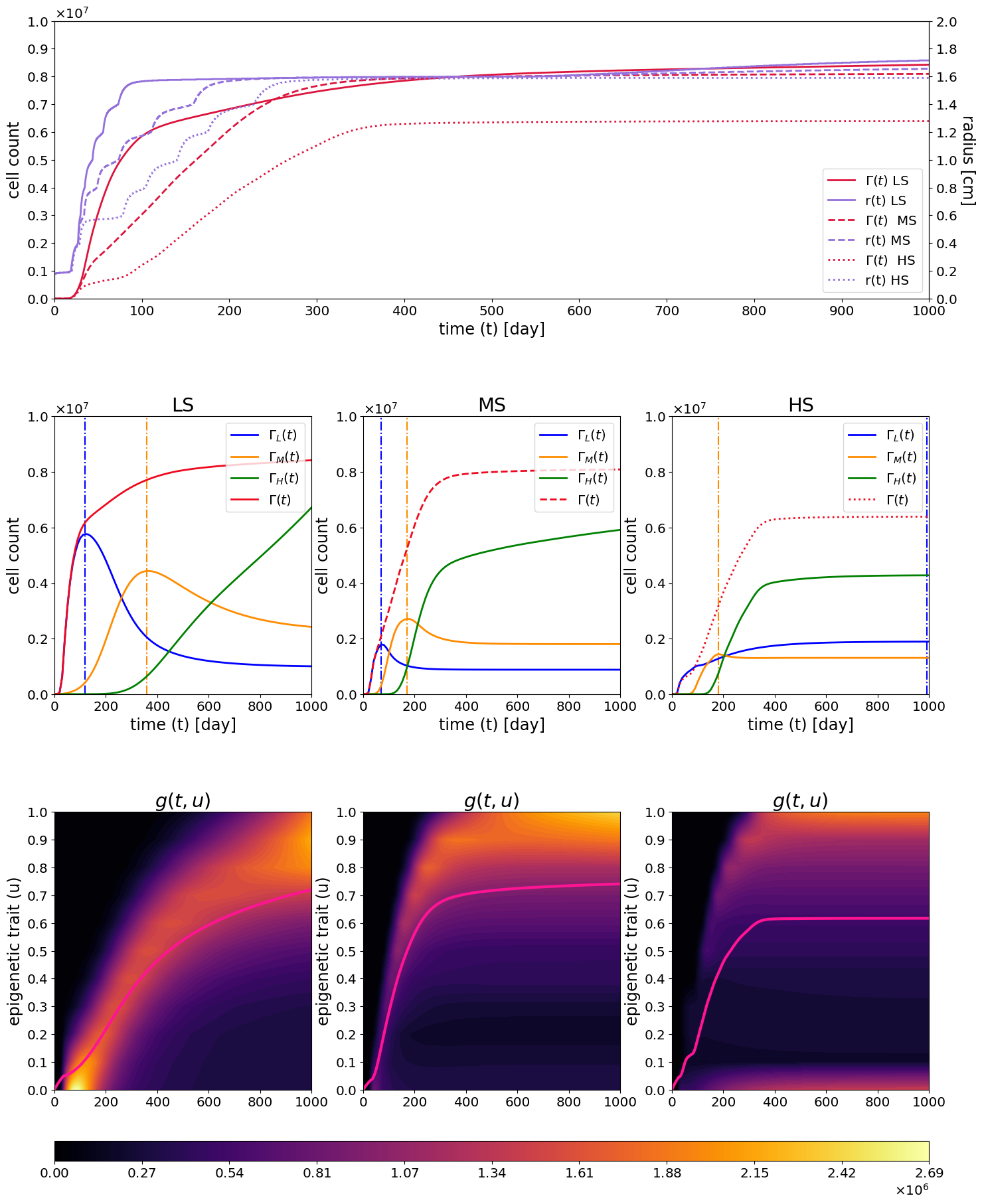}
		\caption{This experiment compares the results from three different simulations in which we vary the intensity of natural selection. In particular, we keep unaltered parameters from the reference simulation, with exception of the selection gradient. We focus on the analysis of a low (LS), medium (MS) and high (HS) selection rate impact, that correspond to $\eta_O=0.1,\, 1,\, 10$ respectively. (a) In first row time evolution of cell count $\Gamma(t)$ and radius $r(t)$ are provided for the three different setting over a time span of $1000$ days. Second and third row are divided in columns according to LS (first), MS (second) and HS (third). (b) Second row shows the evolution in time of band specific and global cell counts $\Gamma(t)$, $\Gamma_L(t)$, $\Gamma_M(t)$, $\Gamma_H(t)$ for $t\in [0,1000]$ days. Vertical lines highlight the times at which cell counts show maximum peaks. (c) Third row provides evolution in time of the epigenetic global density $g(t,u))$ of the population (contourf plot) and highlights the average epigenetic trait $F(t)$ evolution (pink line).}
		\label{fig:selectioncomparison}
	\end{center}
\end{figure}

In the investigation of the effect of the tumour-abiotic factors interaction on the eco-evolutionary story of a mass, it is important to considered that two tumours could have a similar clonal composition at a given point in time, but it does not necessarily indicate that they share similar evolutionary histories and it does not rule out the possibility that their future evolution will diverge significantly, even under the same environmental conditions, \cite{maley2017classifying}. The resulting possible multiple evolutionary pathways  constitute another additional form of variability that turns even harder the challenge to identify the ecological and evolutionary mechanisms that drive the tumour phenotypic evolution.


In this light, we turn on investigating the effects of variations of the selective pressures due to oxygen lack on the evolutionary trajectory of tumour cells and on their consequences on tumour morphology. Selection gradients could be, in this sense, interpreted as measures of how individuals inter-play with the surrounding environment, so a change on their value could be depicted as a variation of the selective pressures on them. 

Therefore, we consider the same set up of the reference simulation ($\Upsilon^F_{SV}$) and we vary the selection parameter $\eta_O$. In particular, we run three experiments in which we consider a low (LS), medium (MS) and high (HS) selection rate, that corresponds to $\eta_O=0.1,\, 1,\, 10$ respectively. The results obtained are summarized in Figure
\ref{fig:selectioncomparison}.

First row of Figure
\ref{fig:selectioncomparison} shows the evolution of the \textit{global cell count} $\Gamma$ and radius $r$ in the three different settings over a time span of $1000$ days. Notably, all cell counts and radii are moving towards the same steady states and the different selection rates just influence the speed and the intermediate interaction dynamics between the cancer cell sub-populations to reach them. This also holds in HS set up in which this tendency is less apparent by the fact that the mass extension proceeds so slowly and the time span considered does not allow to observe it. 
Referring to the evolution of $r$, the selection rate influences both the speed with which the tumour advances in space (the higher is the selection, the lower is the speed) and the invasion mechanism, strongly shaping its profile. 
A steps-like behaviour can be indeed detected that becomes more evident when the selection is higher, coupled with the presence of more extended in time plateaus.

Our results moreover show that a variation of the selection gradient could be also interpreted as a change in the shape of the \textit{Proliferation-Survival trade-off} that affect cancer cells. As shown, a lower selection gradient implies indeed that cells are able to proliferate for a long time also when they are not characterized by the best epigenetic firm which means that, in the case of a lower selection, survival is less expensive, as proved by the higher proliferation/death ratio observed in the dynamics. 

This is strongly reflected in the dynamics of the \textit{global} and \textit{band-specific cell counts}, reported in the second row of Figure \ref{fig:selectioncomparison} comparing the low and the high selection cases. 
In the first column of Figure \ref{fig:selectioncomparison}, referred to the low selection case, the high proliferation/death ratio allows indeed cells characterized by lower epigenetic traits to proliferate fast in the earlier phases, strongly increasing their fraction in the population. When the mass starts to fill all the available space, approaching the carrying capacity of the tissue (as evidenced by the slowdown of the increase in number of the \textit{global cells count} $\Gamma$), selection becomes more predominant, causing their decrease in the population in a sort of \textit{expansion-contraction} dynamics, evidenced by the presence of an high peak in the \textit{low-specific cells count} $\Gamma_L$. The same holds, time-shifted and in a mild manner both with respect the velocity and the amount, in the clonal expansion dynamics of \textit{medium} sub-population $\Gamma_M$, according with the low \textit{proliferating potential} that characterizes these individuals. Conversely, in the last column of Figure \ref{fig:selectioncomparison}, referred to high selection case, no  dynamics of this type are observed, as evidenced by the almost absence of peaks in all the three epigenetic bands which proves that the selection takes place immediately, preventing proliferation in areas of non-optimality and avoiding a significant succeeding decrease in number in low and medium sub-populations. Note that the steady states, in terms of cell count of the bands, correspond to the number of cells present in a tumour that has reached the carrying capacity and that is perfectly selected in all its areas.
In this light, expanding the time-window of observation, the cell count of each band will tend to the same numerical value in all experiments, regardless of the selection rate.

This is reflected in the results presented in the last row of Figure \ref{fig:selectioncomparison} which shows the evolution in time of the \textit{epigenetic global density} $g$ on which the \textit{average epigenetic trait} $F$ is projected in pink. We can indeed recognize in the first column, which refers to the low selection case, the peaks dynamics already observed. In this respect, our results highlight that the previously described \textit{expansion-contraction} phenomenon affect almost all the epigenetic firms with dynamics that differ with respect to the phenotypic trait; time-shifted increasing densities followed by descents affect indeed the entire phenotypic spectrum with slopes that interestingly decrease in accordance to an increasing hypoxia-resistance ability.
An opposite behavior is instead visible in the third column, which refers to the high selection case, where instead the only evident peak is observed for cells characterized by a high hypoxia-resistance which clonal expansion is promoted by the environmental conditions as it can be seen in the third panel of the second row. Notably, we can observe that the strength of the natural selection strongly affect the tumour composition during the mass expansion. Indeed, 
a low natural selection promotes for longer times \textit{coexistence} phenomena, as evidenced by the higher variance of the \textit{epigenetic global density} during the entire simulation. In contrast, an higher selection rate promotes instead quick \textit{out-competition}, forcing tumour cells to promptly acquire the best characteristics and rapidly clonal expanding. 

\begin{figure}[htp!]
	\begin{center}
		\includegraphics[width=0.82\textwidth]{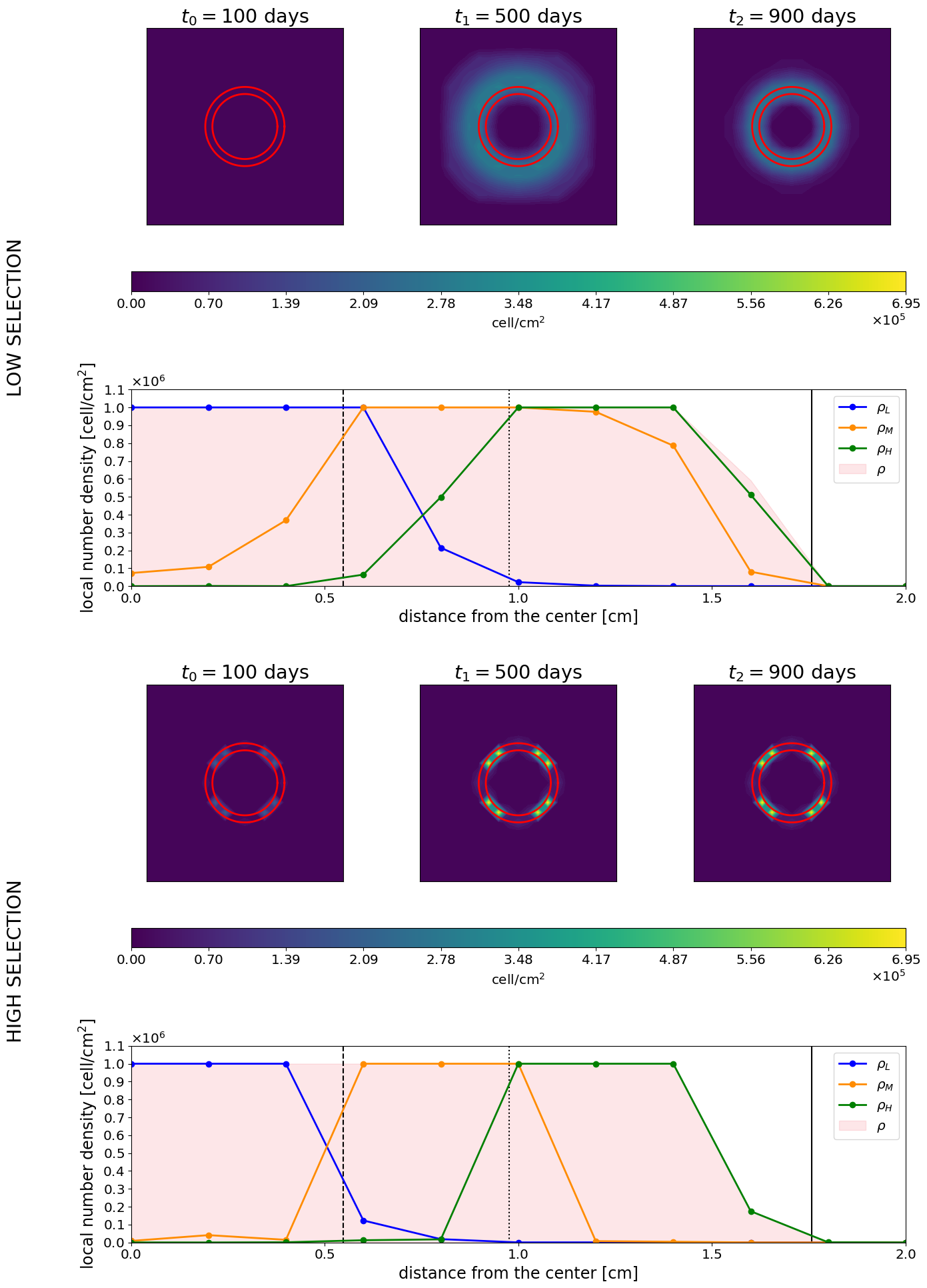}
		\caption{A spatial focused evidence of the dynamic presented in Figure \ref{fig:selectioncomparison} is here shown. Top panels refer to the LS case, the bottom ones to the HS case. The six contourf plots (three for each case) represent the number density $\rho_{0.5}(t,\mathbf{x})$ of viable individuals characterized by the phenotypic trait $u=0.5$ with a variance of $0.05$ at three different times instants ($t=t_1=100$ days, $t=t_2=500$ days, $t=t_3=900$ days).
			Second and fourth rows provide bands-specific number densities $\rho_L(t,\mathbf{x})$, $\rho_M(t,\mathbf{x})$, $\rho_H(t,\mathbf{x})$ at $t= t_F=1000$ days along the segment  $\overline{(0, 0), (2, 0)}$. Vertical lines detect, from left to right, $\Omega_s^L(t_{\textup{F}})$, $\Omega_s^M(t_{\textup{F}})$, $\Omega_s^H(t_{\textup{F}})$, and $\Omega_s^N(t_{\textup{F}})$ (low, medium, and high optimal areas and necrotic area).}
		\label{fig:cerchietti}
	\end{center}
\end{figure}

A spatial focused evidence of this dynamic is represented in Figure \ref{fig:cerchietti}. The graphs refer to the two opposite LS and HS experiments shown in Figure \ref{fig:selectioncomparison}. The top panels refer to the LS case, the bottom ones to the HS case.
The six contour plots (three for each case) represent the number density of viable individuals characterized by the phenotypic trait $u=0.5$, with a variance of $0.05$ at three different times instants ($t=100,500,900$ days) that we denoted with: 
\begin{equation}\label{eq:rho5}
	\rho_{0.5}(t,\mathbf{x})=\int_{0.45}^{0.55} a(t,\mathbf{x},u)\, du.
\end{equation}
The red circles highlight the area of the domain in which, according to the oxygen map, the phenotype $u=0.5$ is the optimal one i.e. they indicate that the tumour mass at the stationary stage should therefore show all the cells characterized by the phenotype $ u = 0.5 $ gathered in these rings. The other plots show the \textit{bands-specific number densities} $\rho_L$, $\rho_M$, and $\rho_H$ at $t=1000$ days along the radius $\overline{(0,0),(2,0)}$.

Coherently with what we observed before, in the LS case, the weak selective pressures allow the tumour to expand in an unregulated way outside the optimal area. Approaching the carrying capacity, the \textit{proliferative potential} slows down and the selective dynamics becomes predominant leading to a progressive death of the cells in areas in which their hypoxia-resistance ability is not sufficiently high to guarantee their survival. 
This dynamics is reflected in the \textit{band-specific cells} plot that shows indeed that the medium band $\rho_M$, in which the chosen phenotype $u=0.5$ belongs, is still largely present also in the areas further away from the vessel where the high epigenetic band is  instead the optimal one,  coherently with the the previous observations that a low selection promotes for longer times \textit{coexistence} phenomena. On the contrary, the invasion dynamics in the HS case are slower but more regulated. The cell density $\rho_{0.5}$ is indeed more spiked and immediately confined to the optimal area. This is reflected also in the \textit{band-specific cell counts} plot that shows that, in the HS case, at the end of the simulation, the presence of each epigenetic band is well localized in the optimality area, coherently with the fact that an high selection rate promotes \textit{out-competition}.

Summarizing, our results communicate the biological notion that the strength of the selective pressures exerted by oxygen,
may shape the emergence of hypoxic resistance in tumours; the way in which such form of resistance
develops depends on the intensity of oxygen-driven selection. In this sense, they moreover suggest that the intensity of the trade-off observed between \textit{proliferation} and \textit{survival} observed in cancer cells plays a significant role in the determination of
future evolution since its shape promote or inhibit \textit{coexistence} or \textit{out-competition} phenomena that completely change tumour composition during its expansion, influencing both its morphology and invasion ability.

\medskip

\section{Discussion}
\label{sec:discussion}
The proposed mathematical modeling approach is able to recapitulate the eco-evolutionary spatial dynamics of tumour cells in their adaptation to hypoxic microenvironments highlighting the impact on this dynamics of the experimentally observed trade-off between \textit{maximizing cell survival} (within the meaning of an increased tolerance to unfavourable conditions) and \textit{maximizing cell growth} that affects cancer cells, \cite{aktipis2013life}. 

Our results confirm the strong influence of the oxygen map on tumour mass development in terms of the areas in which cancer progression takes place and of the differences in both tumour growth speed and epigenetic composition of the population, \cite{ruan2009role}. In this respect, our findings suggest, coherently with experimental observations, that favorable environmental conditions mainly lead to tumours more aggressive in terms of growth and expansion, but detectable in a shorter time, more predictable in behavior and less aggressiveness from the point of view of resistance to treatments. On the other hand, unfavorable environmental conditions mainly slow down tumour growth, but they are closely related to a less predictable course and to more aggressive characteristics in terms of therapeutic resistance, \cite{damaghi2021harsh}. The model outcomes moreover highlight possible mechanisms that underpin this dynamics; on one hand, a slow growth favours indeed a less invasive malignant mass; on the other one, a quiet expansion in terms of both number and extension allows a longer time of tumour non detectability,  during which cancer population could gradually shift its epigenetic dominance towards resistance phenotypes to both hostile environments and treatments. 

The obtained results indicate indeed that tumour masses are strongly affected by heterogeneity in dependence of tumour oxygenation resulting in differences in the geometry of the growing malignant mass. 

In this perspective, our modeling approach is able to capture another important feature of the eco-evolutionary stories of neoplasms: two tumours could indeed have a similar clonal composition at a given point in time but this does not necessarily indicate that they share similar evolutionary histories, and does not rule out the possibility that their future evolution will diverge significantly, even under the same environmental conditions. In particular, we indeed show that, depending on the oxygen selection gradient $\eta_o$, there exist multiple evolutionary pathways that can lead to develop hypoxia-resistance. 

Regarding the role of the \textit{Proliferation-Survival trade-off}, our findings suggest that both its existence and its intensity, which are related to the strength of natural selection, may play an important role in the determination of the phenotypic composition of a cancer population. In particular, we show that its intensity is particularly relevant in shaping the evolutionary trajectory of the individuals and the diversity or variance of trait values among them. In this respect, our results are in line with the classification of the evolutionary and ecological features of neoplasms in \cite{maley2017classifying} identifying the selection gradient $\eta_o$ of the model as a measure of the impact that hypoxia may have on the eco-evolutionary dynamics of tumour cells. Our results show indeed that the strength of the selective pressures exerted by oxygen on tumour cells shapes the emergence of hypoxic resistance in tumours completely changing the pathways in which such form of resistance develops.

The results obtained can be well justified in a niche construction evolutionary perspective. The selection gradients could be indeed interpreted as a measure of how an individual inter-plays with its surrounding environment and this reflects the possibility that if individuals change the way in which they perceive it, the evolutionary trajectory of the population can completely change. In this respect, the traditional view of evolution by Darwinian natural selection may be over-simplistic in the case of cancer development since it sees the environment as an exclusively external
initiator. On the contrary, niche construction theory allows to consider both the natural selection via environmental variables and the alterations of these conditions by the organism itself that result in additional evolutionary or ecological consequences.

\subsection{Perspectives}
\label{subsec:perspectives}

As possible extensions of the model here presented, different aspects can be of particular interest from both the evolutionary and the physical point of view.
In this respect, it would be interesting to consider additional components of tumour environment and their effects on the characterization of the geometric evolution of the tumour mass. This would include: (\textit{i}) the mechanical aspects of the intracellular fluid, (\textit{ii}) the interaction with extracellular matrix in the expansion dynamics, (\textit{iii}) the competition for space and resources with the healthy cells of the host tissue, and (\textit{iv}) the interplay between metabolically active, quiescent and necrotic cells.

With respect to the latter, the necrotic population is already present in the mathematical formulation of our  model, but not particularly investigated. This is because, in this work, we focus on tumour cords, in which the necrotic population develops at the edge of the mass and can, therefore, be reached by macrophages which are responsible for its elimination. On the contrary, in the case of tumour spheroids, the necrotic core develops inside the mass, due to hypoxic conditions induced by the consumption of oxygen by the tumour population itself. In this perspective, it would be interesting to apply the same model to tumour spheroids formation and to consider the effects of the presence of a necrotic core on the morphology and invasion ability of the mass.
Necrotic regions have indeed an active role on the evolutionary dynamics of malignant populations, being responsible for the secretion of cytokines which exert tumour-promoting activity triggering angiogenesis, proliferation, and invasion, \cite{lee2018regulation}. Moreover, in spheroids development, necrotic core extension can be a relevant parameter to be predicted since in multiple experimental works a substantial proportion of necrosis in histopathology samples has been indeed proposed as indicator of tumour aggressiveness associated to poor clinical outcomes, identifying in necrosis extension a valid clinical index to define the tumour degree of advancement and the invasive potential of a growing mass, \cite{pollheimer2010tumor}.

From the evolutionary point of view, it can be of interest to improve our modeling approach including in a more realistic way the trade-offs that affect tumour evolution. 
Biologically, the intensity of a trade-off can be defined as how much, in a population, the development of a determined characteristic is made at the expense of the improvement of an other. 
As reported in literature, it is affected by ecological conditions since the availability of the resources and the hazards of the environment influence how the increase in fitness associated to a change in one trait correlates with a decrease in fitness due to a change in another one. 

In this respect, it could be interesting to introduce an environment-driven dynamical evolution of the intensity of trade-offs to analyze the double influence of (\textit{i}) the epigenetic predisposition and (\textit{ii}) the ecological pressures on the actual behaviour assumed by the cell population. This approach would enable to consider the whole spectrum of trade-off intensity involving both convex and concave profiles, thus improving our results. 

Finally, one of the most important and triggering extension for our model would be to include therapies. The kinetics investigated, as well as the choice of the abiotic factor considered, are indeed able to capture, as done in the simulations, some dynamics that directly impact on therapies efficacy in an early stage tumour, in particular in the case of radiotherapy
In this light, the model presented in this paper laid the groundwork to develop a new mathematical model to investigate how the pre-therapeutic history of a tumour could affect the effectiveness of radiotherapy, how the treatment can be designed to be improved in terms of potency, and how a tumour could evolve in the case of non-eradication (\textit{resistance ability acquisition}), taking into account the impact of environmental geometric characterization and selection forces in the development. This aspect is particularly suitable to be investigated \textit{via} our approach since, in an eco-evolutionary perspective, the emergence of a resistant population can be described in terms of tumour evolution and stems from its intrinsic heterogeneity. All the treatment procedures have indeed a strong impact on our body and act as a environmental stressor on tumour cells, \cite{gatenby2020integrating}. This implies, coherently with the results presented in this paper, that therapeutic agents, inducing modifications of tumour ecology and, consequently, of the fitness landscape of tumour cells, could allow to substantial variations tumours composition. Resistance to therapies reflects, in this sense, the temporal and spatial heterogeneity of the tumour microenvironment as well as the evolutionary potential of cancer phenotypes to adapt to therapeutic perturbations. The treatment resistant hypoxic cells serve indeed as a nidus for subsequent tumour regrowth and repopulation, as well as for regional and distant dissemination, representing a therapeutic dilemma that need to be deeper investigated to guarantee the most effective treatment protocol to possibly avoid relapses.

Furthermore, an extension in this sense of the model could take into account how the effects of radiotherapy doses differ according to the heterogeneity faced at the instant and the location at which the therapy is applied (from both a physical and a phenotypical point of view) investigating how this divergent response could be explicated via niche construction theory. In our eco-evolutionary setting, the experimental evidence of patient-specific response can be indeed justified as mirror of the inter-patients heterogeneity in tumour microenvironment, tumor composition and in the ways in which these two aspects mutual interact. 
Our work can, in this sense, could represent a first step towards the development of a model of radiotherapy which could adapt to patient specific characteristics, in line with the innovative personalized medicine approach, \cite{flashner2019decoding}.

\section*{Conflict of Interest Statement}

The authors declare that the research was conducted in the absence of any commercial or financial relationships that could be construed as a potential conflict of interest.

\section*{Author Contributions}

\noindent\textbf{G. C.} : conceptualization, methodology, software, writing, visualization;\\
\textbf{G. F.} : conceptualization, methodology, writing, visualization;\\
\textbf{M. E. D.} : conceptualization, methodology, writing, visualization, supervision.
\section*{Funding}
This research was partially supported by the Italian Ministry of Education, University and Research (MIUR) through the ``Dipartimenti di Eccellenza'' Programme (2018-2022) -- Dipartimento di Scienze Matematiche ``G. L. Lagrange'', Politecnico di Torino (CUP: E11G18000350001). 

\section*{Acknowledgments}
 All authors are members of GNFM (Gruppo Nazionale per la Fisica Matematica) of INdAM (Istituto Nazionale di Alta Matematica).

\bibliographystyle{ieeetr}
\bibliography{bibliography}
\end{document}